\documentclass[preprintnumbers,amsmath,amssymb,prx,superscriptaddress,twocolumn,a4paper]{revtex4}
\usepackage{balance,doi}
\usepackage[T1]{fontenc}
\usepackage{mathptmx}
\usepackage{graphicx}
\usepackage{color}
\usepackage{amsmath,amssymb}
\usepackage{fixmath}
\usepackage{array}

\begin{document}

\title{Anomalous diffusion for active Brownian particles cross-linked to a networked polymer:\\ Langevin dynamics simulation and theory}
\author{Sungmin Joo}
\affiliation{Department of Physics, POSTECH, Pohang 37673, Republic of Korea}
\author{Xavier Durang}
\affiliation{Department of Physics, POSTECH, Pohang 37673, Republic of Korea}
\author{O-chul Lee}
\email[]{lee572@postech.ac.kr}
\affiliation{Department of Physics, POSTECH, Pohang 37673, Republic of Korea}
\author{Jae-Hyung Jeon}
\email[]{jeonjh@postech.ac.kr}
\affiliation{Department of Physics, POSTECH, Pohang 37673, Republic of Korea}

\date{\today}

\begin{abstract}
	Quantitatively understanding of the dynamics of an active Brownian particle (ABP) interacting with a viscoelastic polymer environment is a scientific challenge. It is intimately related to several interdisciplinary topics such as the microrheology of active colloids in a polymer matrix and the athermal dynamics of the in vivo chromosome or cytoskeletal networks. 
	Based on Langevin dynamics simulation and analytic theory, here we explore such a viscoelastic active system in depth using a star polymer of functionality $f$ with the center cross-linker particle being ABP. We observe that the ABP cross-linker, despite its self-propelled movement, attains an active subdiffusion with the scaling $\langle\Delta \mathbf{R}^2(t)\rangle\sim t^\alpha$ with $\alpha\leq 1/2$, through the viscoelastic feedback from the polymer. Counter-intuitively, the apparent anomaly exponent $\alpha$ becomes smaller as the ABP is driven by a larger propulsion velocity, but is independent of the functionality $f$ or the boundary conditions of the polymer. We set forth an exact theory, and show that the motion of the active cross-linker is a gaussian non-Markovian process characterized by two distinct power-law displacement correlations. At a moderate P{\'e}clet number, it seemingly behaves as fractional Brownian motion with a Hurst exponent $H=\alpha/2$, whereas, at a high P{\'e}clet number, the self-propelled noise in the polymer environment leads to a logarithmic growth of the mean squared displacement ($\sim \ln t$) and a velocity autocorrelation decaying as $-t^{-2}$. We demonstrate that the anomalous diffusion of the active cross-linker is precisely described by a fractional Langevin equation with two distinct random noises.   
\end{abstract}

\maketitle

\section{Introduction}

Brownian particles that disobey the fluctuation-dissipation theorem (FDT)~\cite{zwanzig2001nonequilibrium}--referred to as active Brownian particles (ABPs)--are ubiquitously found in nature~\cite{Geier_ABP,volpe}. Typical examples include the run-and-tumble dynamics of a bacterial microswimmer \cite{bergEColi,Goldstein_PT,Dobnikar}, colloidal particles immersed in a solution of bacteria~\cite{goldsteinPRL2009,wu,yodh}, the motion of Janus particles~\cite{bocquet_ABP}, and the motor-driven directed transport in living cells~\cite{weihs,granick2,jeonmrnp}. In these examples, apart from thermal energy, the Brownian particle takes athermal energy from the environment or from its internal activity, performing self-propelled motion over a certain time scale. Beyond this time scale, the particle undergoes Brownian motion but with an FDT-violated diffusivity. Over the past decade, extensive studies have been conducted for such topics as modeling of the aforementioned ABP systems~\cite{wu,bocquet_ABP,gov2015_PRE,loewen2013_PRE}, their out-of-equilibrium properties~\cite{leonardo2014_PRL,wijland2016_PRL,leonardo2017_SciRep}, and the collective movement of ABPs under various circumstances~\cite{cates2015_review,Bartolo2013,Reverey2015}. 

Regarding the collective dynamics, some efforts were recently made on the so-called active polymer systems, in which a polymer consists of ABPs or is embedded in an active bath embracing active particles; it was investigated how the active fluctuations affect the structure and dynamics of a polymer such as flexibility~\cite{winkler2016_polymers,singh2018_PRE}, swelling~\cite{loewenJCP_2015,zhao2019_polymers,chakrabarti2019_JCP}, looping~\cite{shin2015_NJP}, response to confinement~\cite{CacciutoPRL}, and internal dynamics within the polymer~\cite{gov2014_BPJ,Samanta2016_JPA,osmanovic2017,sakaue2017_actpol,winklerJCP_2017,leticia2011_softmatter}. The active polymer systems may potentially be relevant to modeling and understanding of nonequilibrium biological polymers such as human chromosomes \cite{weber2012_PNAS,gariniNatcomm_2015,PRL2009Telomeres}, microtubule \cite{Gueroui2018_currentbio,Sanchez2012,kawamura2014} or endoplasmic reticulum (ER) network \cite{weissPRE_ER,ER_LIN2014763}, actin \cite{mackintosh2007_science,SonnSegev2017,KosterE1645}, and hydrogels~\cite{Erramilli2005_Biomacromolecules, Wagner2017_biomacromolecules,hydro2019,mucin2019} as well as the transport therein~\cite{elbaum2000_PRL,jeon2011_PRL,mizuno2011_softmatter,Dogic2014_PTRSA,wkkimPRL2019,Seisenberger1929}. 

\begin{figure}
	\centering
	\includegraphics[width=8.5cm]{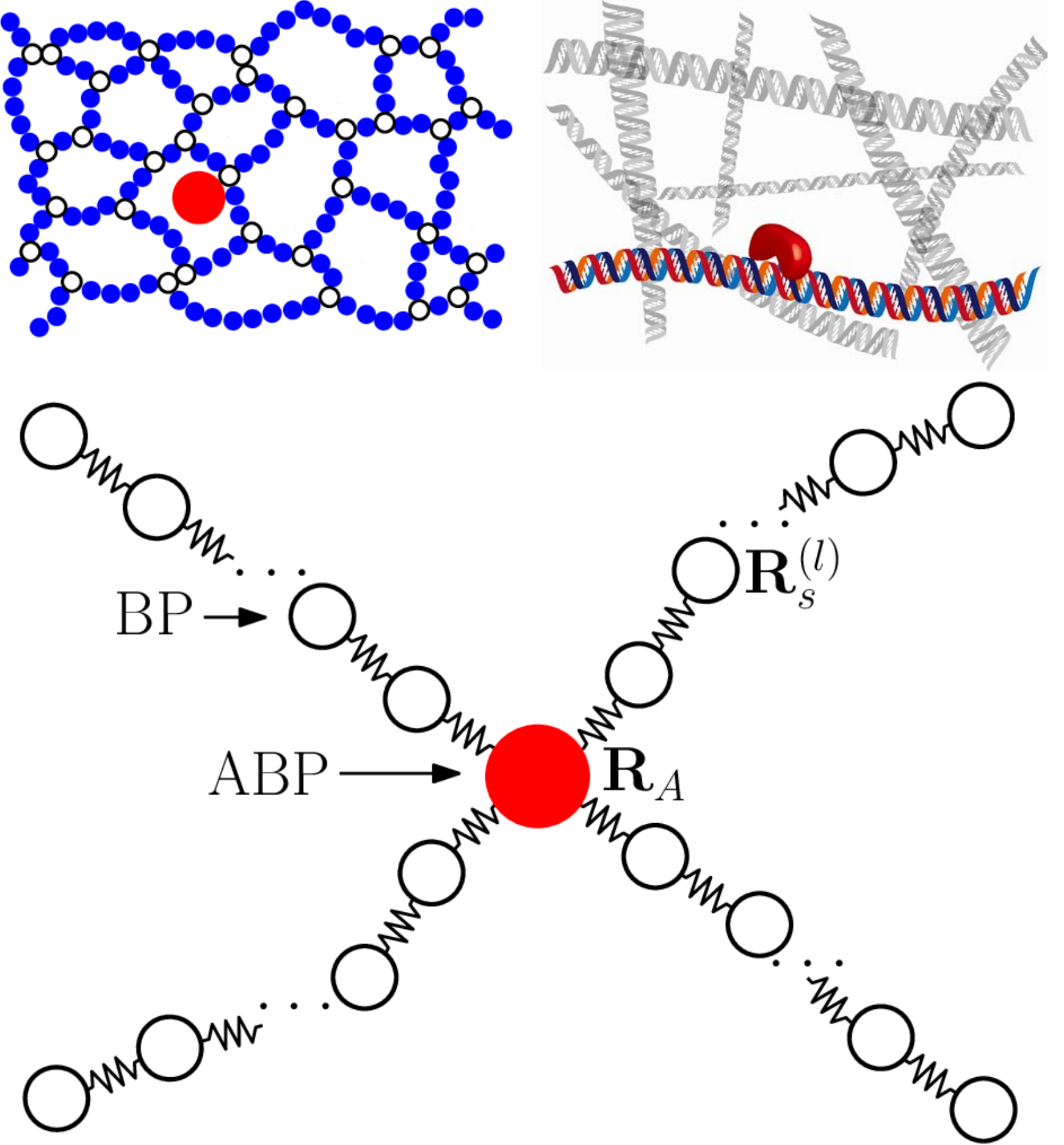}
	\caption{Schematic illustrations showing the active particle in a polymeric environment under consideration. (Top, Left) An active Brownian particle (ABP) is embedded in a cross-linked polymer network. Due to a nonspecific binding interaction between the ABP and the polymer, the particle is allowed to attach onto the network. The ABP has a long-lived negative correlation in its motion through viscoelastic feedback from the meshwork. (Top, Right). \textit{In vivo} biological polymers, such as the chromosome or actin networks, are associated with ATP-consuming macromolecules (red). These active particles strongly bind onto the polymer and locally give rise to nonequilibrium active forces. Through the interactions between the proteins and the polymer, the local segment and protein dynamics can be significantly modified.
		(Bottom) The ABP-polymer composite system as a minimal model for the above two examples. In this model, an ABP at the center (red) is cross-linked with multiple Rouse chains made of ordinary BPs. A similar model was previously considered to model the diffusion of a colloidal particle in a networked polymer~\cite{sprakelPRL}.  
		The number of the connected arms is the functionality $f$ (here, $f=4$) ranging from 2 to 7 in our simulation study. Each arm consists of the same number of BPs $N=100$.  }
	\label{fig1}
\end{figure}

As a related issue in this direction, in this study we are interested in the dynamics of active particles confined to a polymeric environment that embodies viscoelastic features. As exemplified in Fig.~\ref{fig1}, depending on the circumstances, this problem can be viewed as the ABP diffusion confined to a networked polymer [Fig.~\ref{fig1} (Top, Left)] or the local segment dynamics in a bio-polymer driven by the enzymatic activity of an ATP-consuming macromolecule [Fig.~\ref{fig1} (Top, Right)]. Similar problems as to the Brownian particle in a polymeric environment were widely investigated in literature~\cite{schmidt1997_macromol,elbaum2000_PRL,weitz2004_PRL,banksBPJ,weber2010_PRL,jeonNJP2013,wkkimPRL2019,jwshin2020_NatTech,Godec2014}. A number of previous work (including ours) reported that the transport or confined dynamics of a Brownian particle (BP) in a polymer gel often reveals fractional Brownian motion (FBM) \cite{banksBPJ,weber2010_PRL,jeonNJP2013,sprakelPRL}. It is a correlated gaussian process $x_H(t)$ characterized by the autocorrelation $\langle x_H(t)x_H(t')\rangle\propto (|t|^{2H}+|t'|^{2H}-|t-t'|^{2H}) $ and the Hurst exponent $H$ ($0<H<1$)~\cite{mandelbrot}. The mean squared displacement (MSD) of FBM scales as $\langle x_H^2(t)\rangle\sim t^{\alpha}$ with the anomaly exponent $\alpha=2H$ \cite{mandelbrot}. It is known that a tagged monomer in a Rouse chain or a colloidal particle attached to Rouse chains is governed by FBM, with $H=1/4$  ($\langle x^2\rangle\sim t^{1/2}$) \cite{lizanaPRE_fbm,sakauePRE2013,sprakelPRL}, which is called the Rouse dynamics with a Rouse exponent $\alpha=1/2$~\cite{rubinstein2003polymer}. It was shown that BPs diffusing through a polymer network also follow FBM, but with a Hurst exponent varying in the range of $0<H\leq 1/2$, depending on such conditions as the concentration of the polymer and the size of the particle~\cite{banksBPJ,jeonNJP2013}. 

We ask, in this work, how does an ABP connected to a polymeric environment behave in response to the viscoelastic feedback? 
For this study, we consider the minimal networked polymer model initially proposed by Sparkel et al. to study the dynamics of colloidal particles adsorbed onto a networked polymer~\cite{sprakelPRL}. Based on this model, as illustrated in Fig.~\ref{fig1} (Bottom), we construct an ABP-polymer composite system where the ABP cross-links with a set of Rouse chains composed of ordinary Brownian particles. This system mimics either the dynamics of an athermal polymer network driven by active cross-linkers (such as the cross-linker dynamics in an ER network \cite{weissPRE_ER}) or of an active colloid attached to one of the cross-linkers in the network. In the work by Sparkel et al.~\cite{sprakelPRL}, it was shown that when the Brownian colloidal particle is strongly adsorbed onto the networked polymer, the BP displays the Rouse-like dynamics with $\alpha\approx0.5$; if the particle can be easily detached from the polymer, the BP exhibits subdiffusion with $0.5<\alpha\lesssim1$.
Here, we perform Langevin dynamics simulations on our ABP-polymer systems and study how the dynamics of the ABP cross-linker is affected by the viscoelastic interactions with the polymer medium. We find that the ABP cross-linker exhibits apparent anomalous diffusion of $\sim t^\alpha$, where, counter-intuitively, the anomaly exponent $\alpha$ becomes smaller than the Rouse exponent $1/2$ as the active fluctuations become larger. Moreover, the motion of the ABP cross-linker seemingly follows FBM with the Hurst exponent $H=\alpha/2$. The simulation results are then quantitatively elucidated with our analytic theory; mathematically, it is determined that the self-propelled active fluctuations generated by the ABP, eventually, lead to $\sim \ln t$ dynamics in the MSD via the viscoelastic feedback, which turns out to be responsible for the sub-Rouse anomalous diffusion observed in the simulation. Our study gives an insight into the active cross-linker dynamics of networked biofilaments \cite{weiss2017_NJP,ribbeck2018_Natcomm} and reports distinct results compared to the recently studied ABP immersed in a simple polymer solution~\cite{hou2019_softmatter}.

The current paper is organized as follows. In Sec.~\ref{sec:model}, we explain our model for the active Brownian particle and the ABP-polymer system under consideration and then provide the technical information for the Langevin dynamics simulations executed. In Sec.~\ref{sec:simulation}, we provide the computational results from our Langevin simulation. The dynamics of the ABP-polymer systems are extensively investigated upon the change in active fluctuations, the functionality, and the boundary conditions, by analyzing the physical quantities, such as the mean squared displacements, displacement distribution, velocity autocorrelation, velocity cross-correlations, etc. In Sec.~\ref{sec:theory}, we provide our analytic theory based on a simplified ABP-polymer model, where the expressions for the MSD of the ABP cross-linker is derived. The effects of active fluctuations are explicitly studied and tested with the simulation results. Finally, in Sec.~\ref{sec:conclusion}, the main results are summarized with a discussion on the biological implications of this work.

\section{Model \& Simulation Method}\label{sec:model}

In this section, we explain our models for the active Brownian particle and the ABP-polymer systems studied in this work. Next, we provide a detailed description on the Langevin dynamics simulations for our model.

\subsection{Active Brownian particle}
The translational dynamics of an active Brownian particle (ABP) in our model is described by the overdamped Langevin equation~\cite{wu,Samanta2016_JPA,sakaue2017_actpol,um2019}
\begin{equation}\label{eq:abp}
\gamma_A\frac{d\mathbf{R}_A}{dt}=\mathbold{\xi}(t)+\mathbold{\eta}(t).
\end{equation}
Here, $\mathbold\xi$ is the (equilibrium) white gaussian noise from thermal environment satisfying $\langle \xi_\mu \rangle=0$ and $\langle \xi_\mu(t)\xi_{\nu}(t')\rangle=2 \gamma_A k_BT \delta_{\mu,\nu}\delta(t-t')$ [$k_B$: the Boltzmann constant; $T$: the absolute temperature; $\gamma_A$: frictional coefficient; $\mu$ and $\nu$ are the indices of each Cartesian component]; $\mathbold\eta$ represents an active noise from an athermal energy source that leads to the self-propelled motion as well as a breakdown of the FDT~\cite{wu,leonardo2017_SciRep,um2019}. Adopting a  well-known model in literature~\cite{wu,leonardo2017_SciRep,um2019}, here, we model the active noise as an Ornstein-Ulhlenbeck (OU) noise, which has zero mean and an exponentially decaying autocorrelation of the form:
\begin{equation}\label{eq:OUauto}
\langle \eta_\mu(t)\eta_{\nu}(t')\rangle=\frac{\gamma_A^2  v_\mathrm{p}^2 }{3}\delta_{\mu,\nu}\exp{(-|t-t'|/\tau_A)} .
\end{equation}
In this expression,  $\tau_A$ is the correlation time of the noise and  $v_\mathrm{p}$ is the propulsion velocity quantifying the degree of self-propelled mobility. The average propulsion velocity from the active noise becomes $v_\mathrm{p}$ at the noise strength defined above (Fig.~S1 and Sec.~II A in the ESI\dag). In this study, we alternatively re-express $v_\mathrm{p}$ in terms of the P{\'e}clet number $\mathrm{Pe}$, i.e. the ratio of the active diffusion rate to the thermal diffusion rate. It is defined as \cite{volpe} 
\begin{equation}
\mathrm{Pe}=\frac{v_\mathrm{p} \sigma}{D}
\end{equation}
where $\sigma$ is the diameter of the particle and $D=k_BT/\gamma$ is its thermal diffusivity. The above OU active noise (in each Cartesian component) is governed by the Langevin equation \cite{winkler2016_polymers,um2019}
\begin{equation}\label{eq:vp}
\frac{d\eta}{dt}=-\tau_A^{-1}\eta(t) +\sqrt{\frac{2\gamma_A^2 v_\mathrm{p}^2}{3\tau_A}}\zeta(t)
\end{equation}
where $\zeta$ is a white gaussian noise satisfying $\langle\zeta\rangle=0$ and $\langle\zeta(t)\zeta(t')\rangle=\delta(t-t')$. See Fig.~S1 and Sec.~II A (ESI\dag) for the autocorrelation of the OU active noise generated by Eq.~\eqref{eq:vp}. 

\subsection{ABP cross-linked to a Rouse networked polymer}
Using the ABP cross-linker and the ordinary Brownian beads, we set up an active polymer system, as illustrated in Fig.~\ref{fig1}. Several Rouse polymers of the same length are cross-linked with an ABP at the center. With the exception of this cross-linker particle, all beads are the ordinary BPs. This star polymer can be understood as a unit cell of a polymer network having the functionality $f$ (the number of connected chains). For simplicity, the size of the ABP is assumed to be the same as the other beads in the polymer. The dynamics of this active polymer is described by the following Langevin equations: 
\begin{equation}\label{eq:method_abp}
\begin{aligned}
\gamma_A \frac{d\mathbf{R}_{A}}{dt}&=-k\sum_{l=1}^{f}(\mathbf{R}_{A}-\mathbf{R}_{1}^{(l)})+\mathbold\xi(t)+\mathbold{\eta}(t), \\
\gamma\frac{d\mathbf{R}_{s}^{(l)}}{dt}&=-k(2\mathbf{R}_{s}^{(l)}-\mathbf{R}_{s+1}^{(l)}-\mathbf{R}_{s-1}^{(l)})+\mathbold{\xi}_{s}^{(l)}(t). 
\end{aligned}
\end{equation}
The first equation describes the motion of the ABP cross-linker represented by $\mathbf{R}_A(t)$ [$\equiv\mathbf{R}_0(t)$ with the index $0$ denoting the center bead]; the vector $\mathbf{R}_{s}^{(l)}(t)$ denotes the position of $s$-th monomer  in $l$-th linear chain (where $s\in \{1, 2, \ldots, N\}$ and $l\in \{1, 2,\ldots,f\}$) and $k$ is the spring constant for the harmonic potential between neighboring beads. $\mathbold{\eta}$ is the active OU noise introduced in Eqs.~\eqref{eq:abp} \& \eqref{eq:OUauto} for the self-propelled motion of the ABP. The second equation describes the dynamics of the Rouse chains in the polymer network. $\mathbold{\xi}$ and $\mathbold{\xi}_s^{(l)}$ are the $\delta$-correlated thermal noises for the cross-linker and the remaining particles, respectively, which are independent of one another and have a variance of $2\gamma k_BT$ for each Cartesian component. 

Two distinct boundary conditions for the arms are considered in our study. The first case is the pinned arms where the last $N$-th beads in the arms are fixed in space (i.e., $d \mathbf{R}_{N}^{(l)}/dt=0$) as shown in Fig.~\ref{fig1}. The other case is the free boundary condition where the end beads move freely ($\frac{\partial \mathbf{R}_s^{(l)}}{\partial s}|_{N}=0$).  Based on our simulations for both cases, we show that the monomer dynamics at the time scales of our interest are hardly affected by the boundary conditions, which merely determine the long-time limit behavior of whether the system has a confined motion or free diffusion. By virtue of this, we mostly focus on the case of pinned arms in our simulation study and discuss the boundary effects at the end of the simulation section. In the current study, we simulate the ABP-polymer systems of functionality $f$ ranging from 2 to 7, where each arm has $N=100$.  

\subsection{Langevin Dynamics Simulations}

We perform the Langevin dynamics simulation for our model based on the equation of motion~\eqref{eq:method_abp}.  In the simulation, we use the basic units for length $[\sigma]=b$ (the bead diameter), time $[\tau]=1$~ns, and energy $[\epsilon]=k_BT$ ($T=300$~K). Using these units we solve the Langevin equation~\eqref{eq:method_abp} within the scheme of the second-order Runge-Kutta algorithm with the Heun method~\cite{hellevik2018numerical}. 
Further information about the simulation scheme and the boundary condition is provided in the ESI\dag.

In our Langevin simulations we chose:  the diffusivity of (A)BP beads $D=k_BT/\gamma=2$ $[\sigma^2/\tau]$,  the spring constant $k=3k_BT/b^2=3$ $[\epsilon/\sigma^2]$, the integration time $\Delta t=0.01 \tau$, and the total simulation time ranging from $\mathrm{T}=2\times10^3 \tau$ to $\mathrm{T}=2\times10^5 \tau$. We also set $\tau_A=0.083~\tau$, and the propulsion velocity takes the values $v_\mathrm{p}=0$ (BP limit), 10, 20, 30, 50, and 100. The corresponding  P{\'e}clet numbers for each $v_\mathrm{p}$ are simply $\mathrm{Pe}=v_\mathrm{p}/2$ for our given choice of parameters. For simplicity, the frictional coefficient of a bead in the polymer network is set to be $\gamma=\gamma_A$. For a given parameter set, 100 samples are run for statistics.

Prior to our simulation on the ABP-polymer systems, we have validated our code by simulating some model systems that an available corresponding analytic theory: In Fig.~S1, a single ABP in free space has been simulated for various $\mathrm{Pe}$ and their excellent agreement with the exact analytic theory has been demonstrated. Additionally, in Sec.~\ref{sec:ABP} [Fig.~\ref{fig9}], the comparison between our simulation results and the analytic theory of the ABP-polymer system is provided (including the limit of the Rouse polymer at $\mathrm{Pe}=0$). 

\section{Results: Langevin dynamics simulations}
\label{sec:simulation}

\subsection{Anomalous diffusion of the ABP cross-linker}

To begin with, let us study how the dynamics of an ABP is altered when cross-linked to a networked polymer. We have simulated the ABP-polymer systems with three arms ($f=3$) at varying P{\'e}clet numbers [see Fig.~\ref{fig1}(Bottom)]. With these simulation data, we measure the mean squared displacement of the ABP cross-linker by the definition $\langle \Delta \mathbf{R}^2_A(t)\rangle\equiv \langle \overline{ [\mathbf{R}_A(t_0+t)-\mathbf{R}_A(t_0)]^2}\rangle_\mathrm{sp}$ where $ \overline{ \Delta\mathbf{R}^2_A(t; t_0)}$ represents a time-averaged (TA) MSD from a single trajectory and $\langle \cdot \rangle_\mathrm{sp}$ is the ensemble-averaging of TA MSDs over $100$ samples simulated at given parameters~\cite{ralfpccp2}. 
\begin{figure}
	\centering
	\includegraphics[width=9 cm]{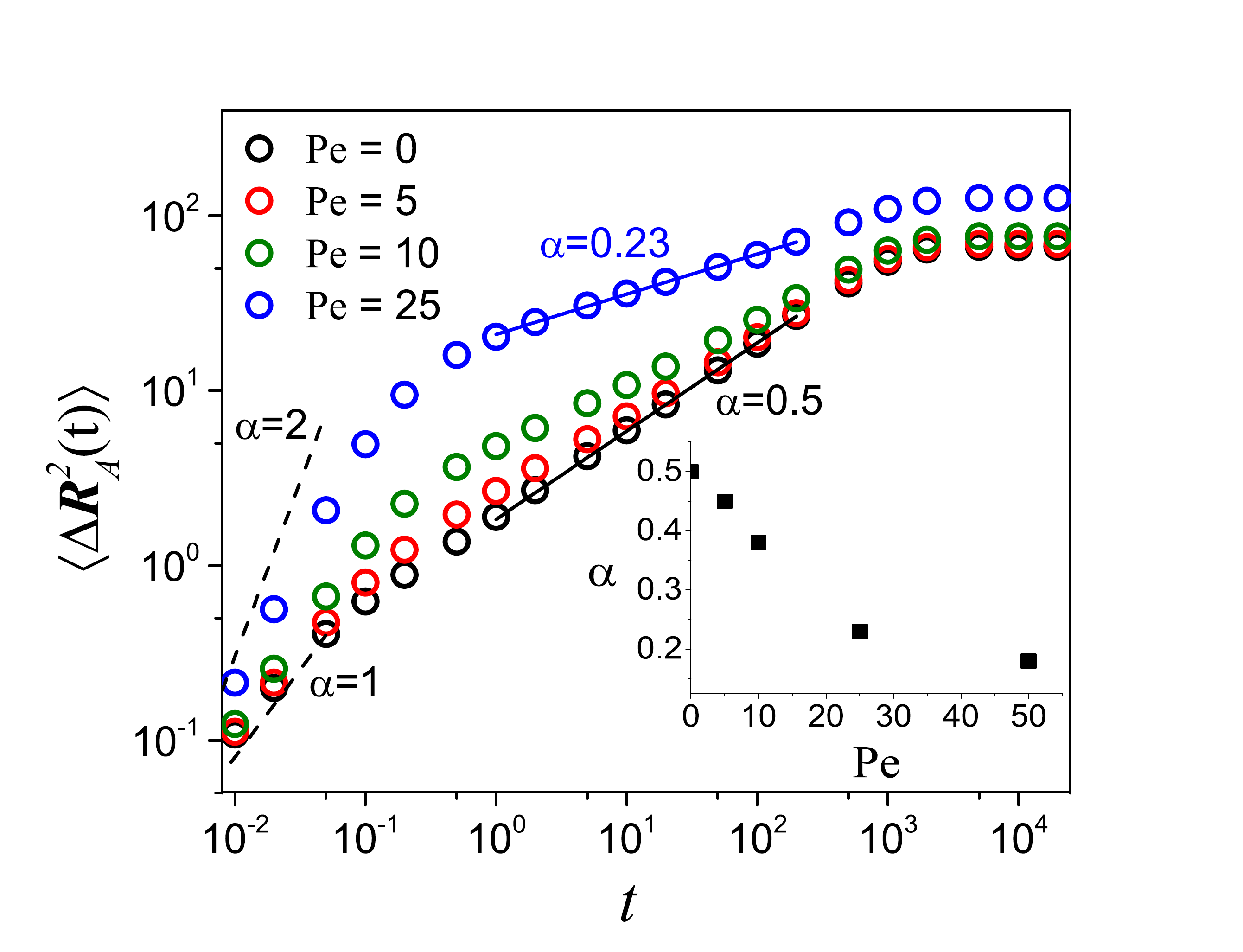}
	\caption{Mean squared displacements (MSDs) of the ABP cross-linker in a networked polymer when varying the P{\'e}clet number ($\mathrm{Pe} = v_\mathrm{p}/2$). From Bottom to Top, $\mathrm{Pe}$ is zero (the BP limit), 5, 10, and 25. The apparent MSD exponent (solid lines) in the Rouse regime ($\sim 10^{0}\ldots 10^2$ $\tau$ ) is annotated for $\mathrm{Pe}=0$ and $25$. The inset shows the fitted $\alpha$ values from MSD $\sim t^\alpha$ in this regime as a function of $\mathrm{Pe}$. Two dashed lines denote the guides for the Fickian and the ballistic limits. In the plot, the functionality is $f=3$ and the end beads of each arm pinned in space. Here (and in the subsequent similar plots below), the error bar is smaller than the symbol size. The basic units in $x$- and $y$-axes are $[\tau]$  and $[\sigma^2]$, respectively. }
	\label{fgr:msd}
\end{figure}
In Fig.~\ref{fgr:msd}, we plot the MSDs of the ABP cross-linker at varying $\mathrm{Pe}(=v_\mathrm{p}/2)$. First, at the BP limit of $\mathrm{Pe}=0$, the MSD exhibits the expected result reported in the literature~\cite{sprakelPRL}. The Brownian cross-linker, followed by the short-time Fickian dynamics, crosses over to subdiffusive dynamics with the anomaly exponent $\alpha=1/2$. This is referred to as the Rouse dynamics, resulting from the collective motion of a polymer network that persists, approximately, up to the Rouse relaxation time $\tau_R=4\gamma N^2/(\pi^2 k)\sim 670$. Beyond this time the BP cross-linker shows confined motion because of the fixed boundary condition. For the ABP cross-linkers, the motion is super-diffusive (or ballistic) at $t\lesssim \tau_A$, during which the ABP motion is directional due to the active noise. For $\tau_A \lesssim t\lesssim \tau_R$, the ABP shows a transition from superdiffusion to subdiffusion with the anomaly exponent $\alpha<1/2$. The inset shows the plot of $\alpha(\mathrm{Pe})$ obtained by a fit with MSD $\propto t^\alpha$ over the time window $[10^0,10^2]$. Surprisingly, the result demonstrates that the ABP motion is more subdiffusive (with smaller $\alpha$) as $\mathrm{Pe}$ is increased. It seems that the net effect of the active noise is counter-intuitive when the ABP attains the long-lived negative feedback from the viscoelastic environment. Importantly, we point out that such an unexpected slower subdiffusion can occur in situations where a single ABP is connected to a passive polymer system. If the polymer system consists of the same ABPs, we find that the ABP cross-linker always displays a Rouse-like subdiffusion of $\alpha=1/2$, regardless of $\mathrm{Pe}$ (Fig.~S2). This is also consistent with the previous work~\cite{osmanovic2017}. 
After the Rouse relaxation time $\tau_R$ the MSD reaches a plateau, whose amplitude increases for higher $\mathrm{Pe}$.  

\begin{figure}[]
	\centering
	\includegraphics[width=8.5cm]{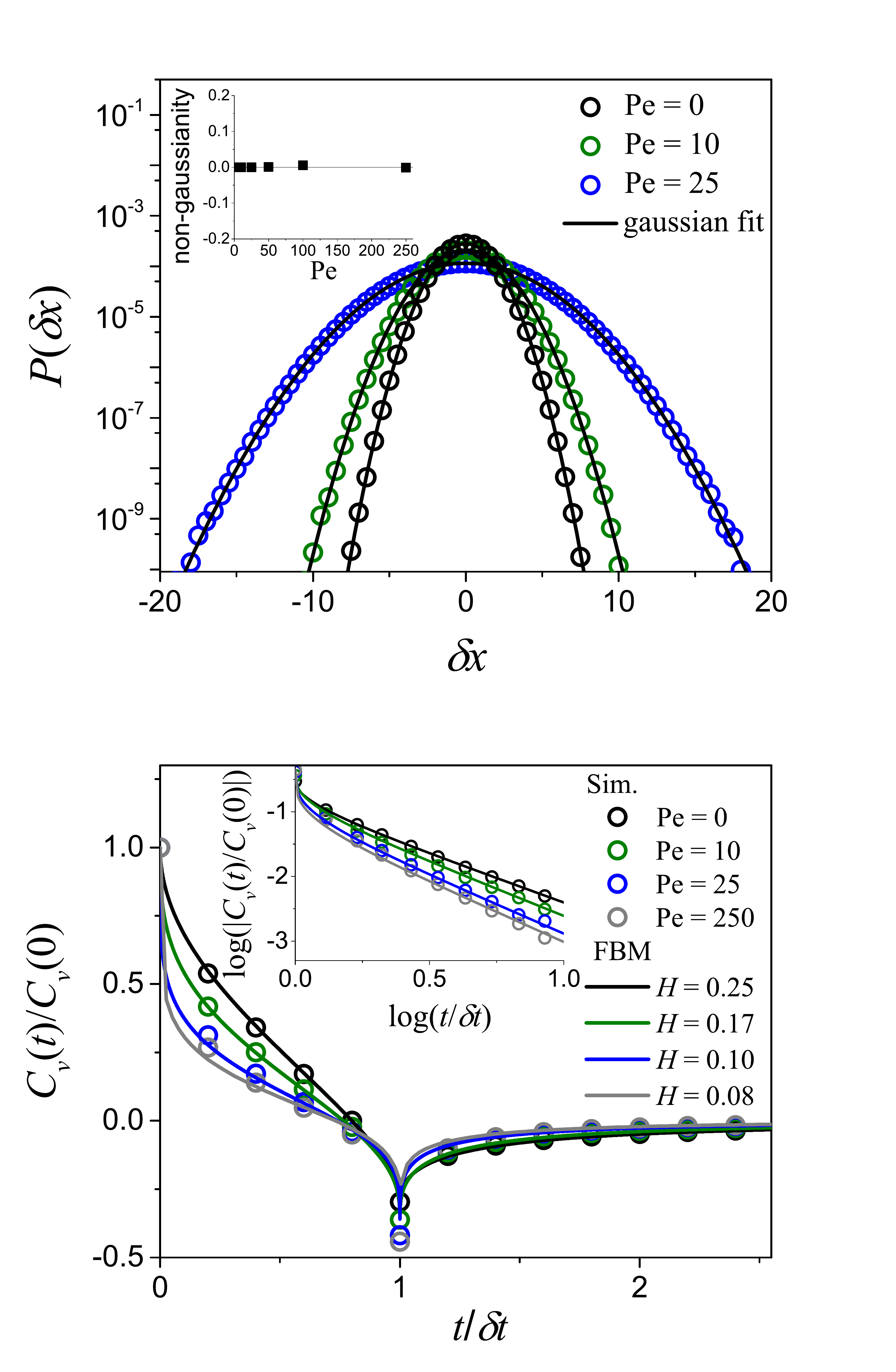}
	\caption{Stochastic nature of the ABP cross-linker dynamics. (Top) Displacement distribution $P(\delta x|t)$ of the ABP cross-linker for a given lag time $t=10$. The $P(\delta x)$s are all normalized to unity. The solid lines are the gaussian fits to the data. Inset: The non-gaussian parameter $\langle (\delta x)^4\rangle / [3\langle (\delta x)^2\rangle^2]-1$ vs. $\mathrm{Pe}$. (Bottom) Velocity auto-correlation functions (VACFs) of the ABP cross-linker for $\mathrm{Pe} = 0$ (the BP limit), 5, 10, and 25. The definition of the velocity and its normalized VACF are given in the main text. The time interval $\delta t$ for defining the velocity is $\delta t=10~\tau$. The solid lines are Eq.~\eqref{eq:vacf_fbm}, which are the normalized VACF curves for FBM with the Hurst exponent $H(=\alpha/2)=0.25$, 0.23, 0.19, and 0.12. Inset: the tail parts of the same VACF curves in the main figure for $t/\delta t>1$ in the log-log scale. }
	\label{fgr:dist}
\end{figure}

For the BP cross-linker (at $v_\mathrm{p}=0$), it is well-known that the observed subdiffusive motion of $\alpha=1/2$ belongs to fractional Brownian motion (FBM) with the Hurst exponent $H(=\alpha/2)=1/4$ \cite{mandelbrot, lizanaPRE_fbm,sakauePRE2013}. In this case, the cross-linker motion is gaussian and has an algebraically decaying negative correlation in its velocity, such that $\langle \mathbf{v}(t_1)\cdot \mathbf{v}(t_2)\rangle\sim \alpha(\alpha-1)|t_1-t_2|^{\alpha-2}$~\cite{ralfpccp2}. 

We scrutinize the stochastic nature of the ABP cross-linkers. First, the gaussianity of the process is examined by plotting the displacement distribution (i.e., the Van-Hove self-correlation function) $P(\delta x)$ with a gaussian fit. Fig.~\ref{fgr:dist} (Top) shows $P(\delta x|t=10)$ and its best gaussian fit for various $\mathrm{Pe}$ conditions. The analysis suggests that $P(\delta x)$ follows a gaussian distribution. To confirm that $P(\delta x)$ is indeed absence of a non-gaussian tail originating from the large displacements from the active motion, especially, at high P{\'e}clet numbers, we measure the non-gaussian parameter $\langle (\delta x)^4\rangle / [3\langle (\delta x)^2\rangle^2]-1$ for increasing $\mathrm{Pe}$ up to $250$ [see the inset]. This demonstrates that, regardless of $\mathrm{Pe}$, the non-gaussian parameter is as nearly zero as the genuine gaussian displacement is when realized by a gaussian random number generator. Based on these analyses, we conclude that the anomalous diffusion of the ABP cross-linkers is gaussian. 

Next, we study the directional memory in the motion of the ABP cross-linker via the velocity autocorrelation function (VACF). Here, the average velocity is defined from a trajectory over an arbitrary time interval $\delta t$ by $\mathbf{v}(t;\delta t)=[\mathbf{R}(t+\delta t)-\mathbf{R}(t)]/\delta t$. Then, VACF is given by $C_v(t;\delta t)=\langle \mathbf{v}(t+t_0;\delta t)\cdot \mathbf{v}(t_0;\delta t)\rangle$. The normalized VACFs are plotted in Fig.~\ref{fgr:dist} (Bottom) for several $\mathrm{Pe}$ conditions. The inset shows the tail part of the same VACFs in the log-log scale. We compare the data with the normalized  VACF curves of FBM~\cite{ralfpccp2,jeonPRL_lipid}
\begin{equation}\label{eq:vacf_fbm}
C_v(t;\delta t)=\frac{(t+\delta t)^{2H}-2t^{2H}+|t-\delta t|^{2H}}{2\delta t^{2H}}
\end{equation}
for a process with a Hurst exponent $H=\alpha/2$ and time interval $\delta t$. The comparison shows that the VACF of the ABP cross-linker is hardly distinguished from that of FBM (including the tail part shown in the inset) for $\mathrm{Pe}\lesssim 25$. The excellent agreement strongly suggests that the anomalous diffusion of the ABP cross-linkers (with the apparent scaling of MSD$\sim t^\alpha$) behaves as FBM with a Hurst exponent $H=\alpha(\mathrm{Pe})/2$. Eq.~\eqref{eq:vacf_fbm} then indicates that the ABP cross-linker has a power-law decaying anticorrelation in its displacement as VACF$\sim \alpha(\alpha-1)t^{\alpha-2}$. We point out that such an FBM-like ABP dynamics fails to observe if $\mathrm{Pe}\gtrsim 25$. As an extreme case, the VACF at $\mathrm{Pe}=250$ is plotted in the figure. The curve does not exactly follow the VACF of FBM [Eq.~\eqref{eq:vacf_fbm}] at both short and large times, although the deviation appears to not be significant. We clarify the stochastic identity of the ABP cross-linker in the section of analytic theory (Sec.~\ref{sec:ABP}).        

\subsection{The polymer dynamics in the presence of the ABP cross-linker }
\begin{figure}
	\centering
	\includegraphics[width=8.5cm]{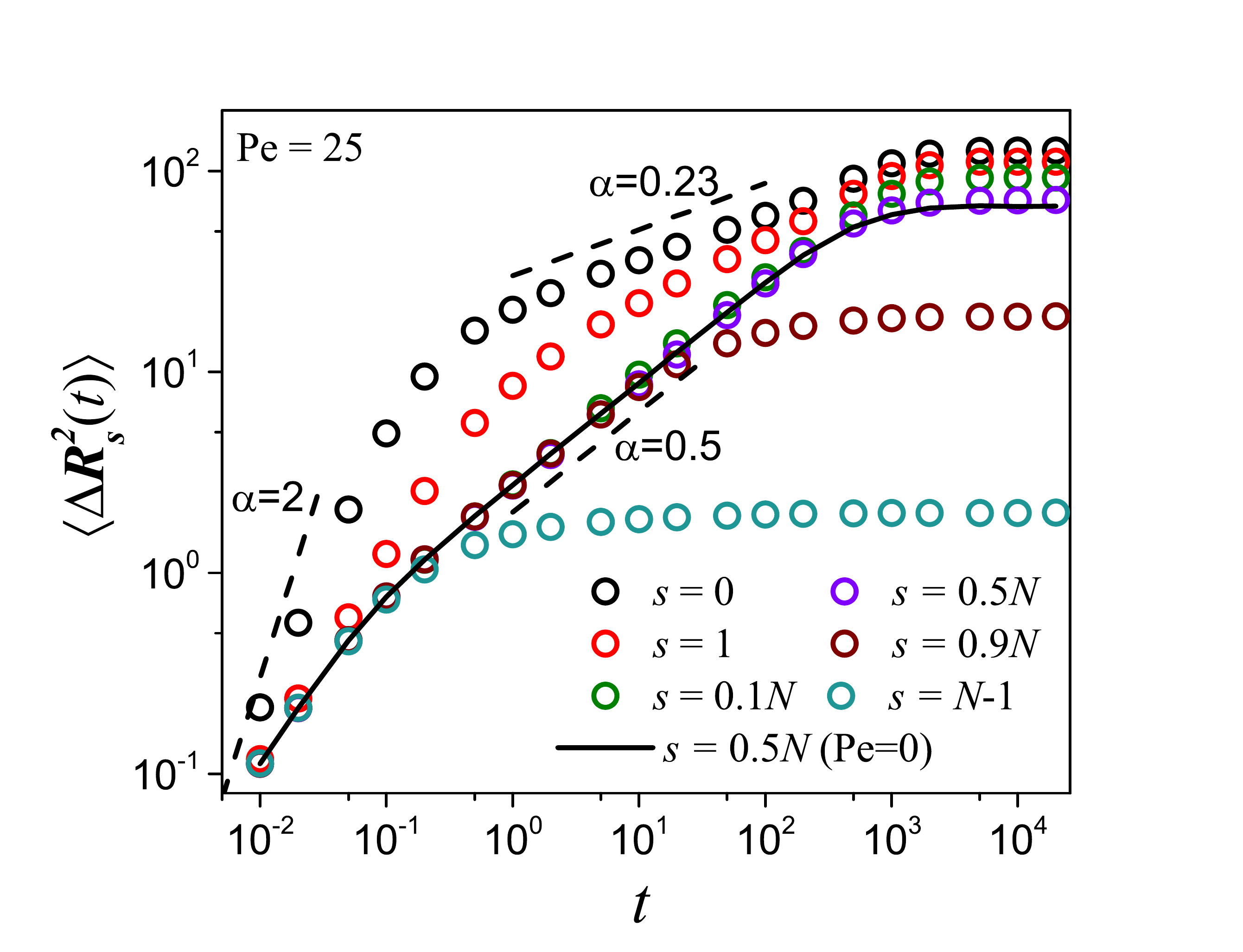}
	\caption{MSDs for the passive beads in the arm chains for varying $s$ from the center cross-linker to the end. From Top to Bottom, the curve shows the particle at $s=0$ (the ABP cross-linker), 1, $0.1N$, $0.5N$, $0.9N$, and $N-1$ (here, $N=100$). The solid line represents the MSD from the simulation for $s=0.5N$ at $\mathrm{Pe}=0$. We use the functionality $ f=3$ and $\mathrm{Pe}=25$. }
	\label{fgr:monomer}
\end{figure}

We now turn to the dynamics of the beads in the arms and study how their collective dynamics--the Rouse dynamics at $\mathrm{Pe}=0$--are modified by the presence of the ABP. In Fig.~\ref{fgr:monomer}, we plot the MSDs, $\langle \Delta\mathbf{R}^2_s(t)\rangle=(1/f)\sum_{l=1}^f \langle \overline{[\mathbf{R}_s^{(l)}(t+t_0)-\mathbf{R}_s^{(l)}(t_0)]^2}\rangle_\mathrm{sp}$, for the $s$-th beads in the arm chains (at a fixed $\mathrm{Pe}=25$). The six curves from Top to Bottom correspond to $s=0$ (the ABP cross-linker), 1, $0.1N$, $0.5N$, $0.9N$, and $N-1$ [$N=100$]. As a reference, on top of these we plot the MSD for $s=N/2$ at $\mathrm{Pe}=0$ (solid line). From the figure we find the following interesting features for the athermal polymer dynamics: (1) In the Rouse regime the Brownian beads exhibit subdiffusion where the anomaly exponent $\alpha$ becomes larger with an increasing $s$ and, concurrently, the amplitude is significantly smaller. (2) At the middle of the arm, $s=N/2$, the MSD curve collapses onto the reference curve (solid line) at  $\mathrm{Pe}=0$,  having the maximum $\alpha=1/2$ (the Rouse exponent). Interestingly, the active fluctuations at the linker ($s=0$) almost decay out around the middle of the arm, thus, the motion of the bead is thermal. (3) For the beads where $s>N/2$, the effect of the active noise is negligible; the particles exhibit a Rouse dynamics under confinement. 

\begin{figure}
	\centering
	\includegraphics[height=8.5cm]{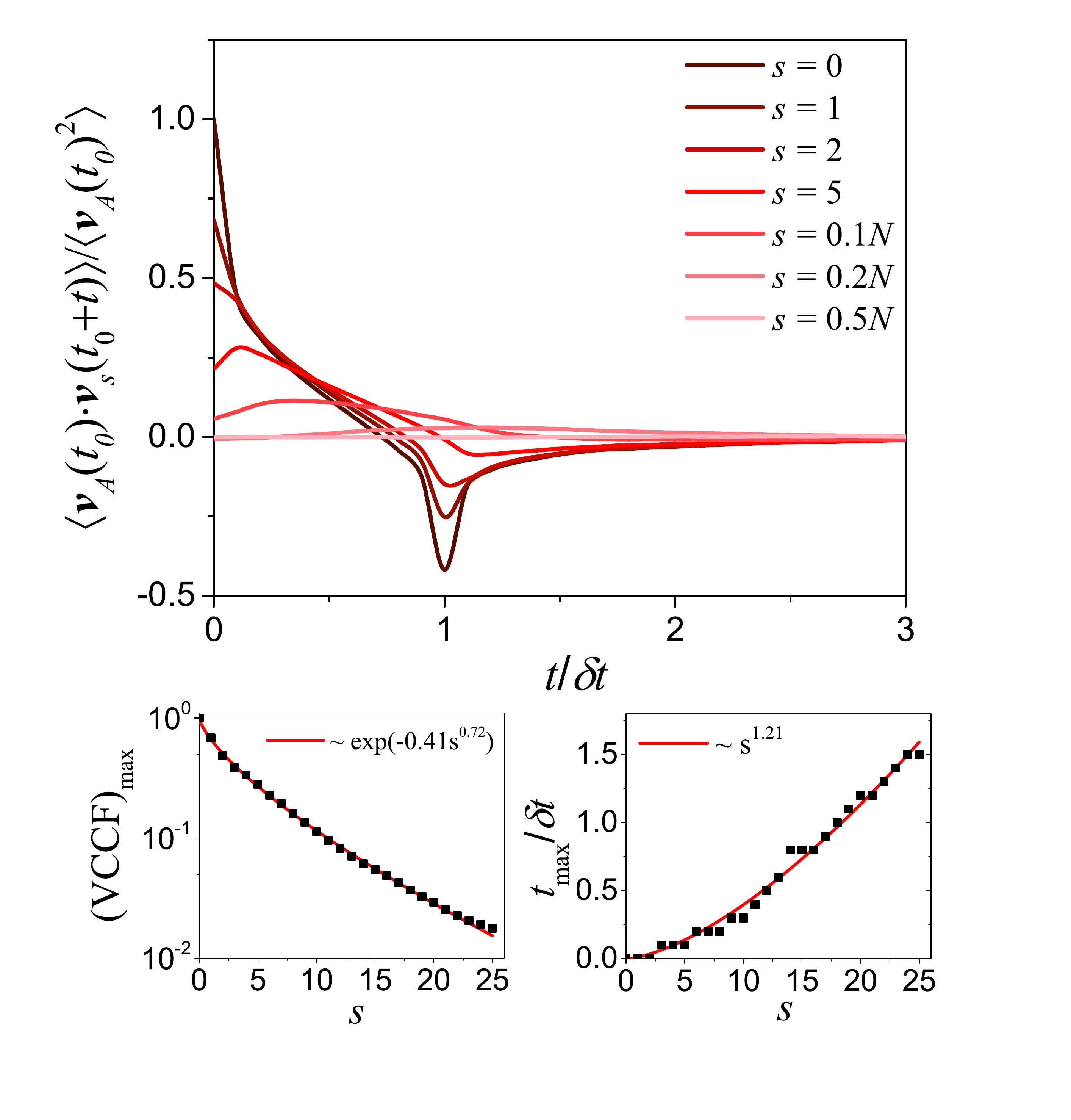}
	\caption{Velocity cross-correlation function (VCCF) between the ABP cross-linker and the BPs at various distances $s$. VCCF is defined as $\langle \mathbf{v}_s(t+t_0;\delta t) \cdot \mathbf{v}_A(t_0;\delta t)\rangle$ where the velocities are defined in the same manner as in VACF (Fig.~\ref{fgr:dist}) and the time interval $\delta t=10~\tau$.  (Top) VCCFs as a function of lag time $t$. The results for the systems with $f=3$ and $\mathrm{Pe}=25$. (Bottom, Left) The maximal value of the normalized VCCFs as a function of the bead index $s$. The red solid line depicts the stretched exponential fit to the data. (Bottom, Right) The time for the max VCCF, $t_\mathrm{max}/\delta t $, as a function of the bead index $s$. The red solid line is the power-law fit to the data. }
	\label{fgr:vcc}
\end{figure}

To examine the impact of the active fluctuations along the chain, we measure the velocity cross-correlation function (VCCF) between the ABP ($s=0$) and the BPs at $s\geq1$. Plotted in Fig.~\ref{fgr:vcc} are the VCCF curves normalized by $\langle \mathbf{v}_A^2\rangle$. It shows that for small $s$ the VCCFs have a negative dip at $t\sim \delta t$ ($\delta t=10$ in the plot), suggesting that the motion of these BPs is anti-correlated with that of the ABP cross-linker. Here we find that the degree of anti-correlation  decreases with $s$ while the initial positive-correlation lasts longer; for instance, the bead at $s=N/5$ almost has a positive-correlation with the ABP without anti-correlation over time. By a numerical analysis in Fig.~\ref{fgr:vcc} (Bottom, Left) we find that the positive correlation (i.e., the maximal value of VCCF) decays with $s$ in a stretched exponential manner of $\sim \exp(-As^{0.72})$, where the stretched exponent varies depending on $\mathrm{Pe}$. For $s\gtrsim N/2$ the cross-correlation is found to be almost zero. This is consistent with our observation in the MSDs, where the beads after the half-length arm are hardly affected by the active fluctuations at the center. We also learn that the time for the highest positive correlation increases with $s$. This behavior can be understood as a time delay for the active fluctuations occurring at the center to arrive at a distant bead. The subplot in Fig.~\ref{fgr:vcc} (Bottom, Right) shows that the peak time $t_\mathrm{max}$ grows algebraically with $s$ as $\sim s^{1.21}$. However, the growth exponent 1.21 is not universal and tends to decrease with $\mathrm{Pe}$. This behavior can be viewed such that the number of beads affected by the active noise grows with time as  $s\sim t_\mathrm{max}^{1/1.21}$ and  the corresponding exponent increases with $\mathrm{Pe}$ (the exponent is $1/2.47$ at $\mathrm{Pe}=0$).

\subsection{The effect of functionality ($f$)}
\label{subsec:f}
\begin{figure}
	\centering
	\includegraphics[width=8.5cm] {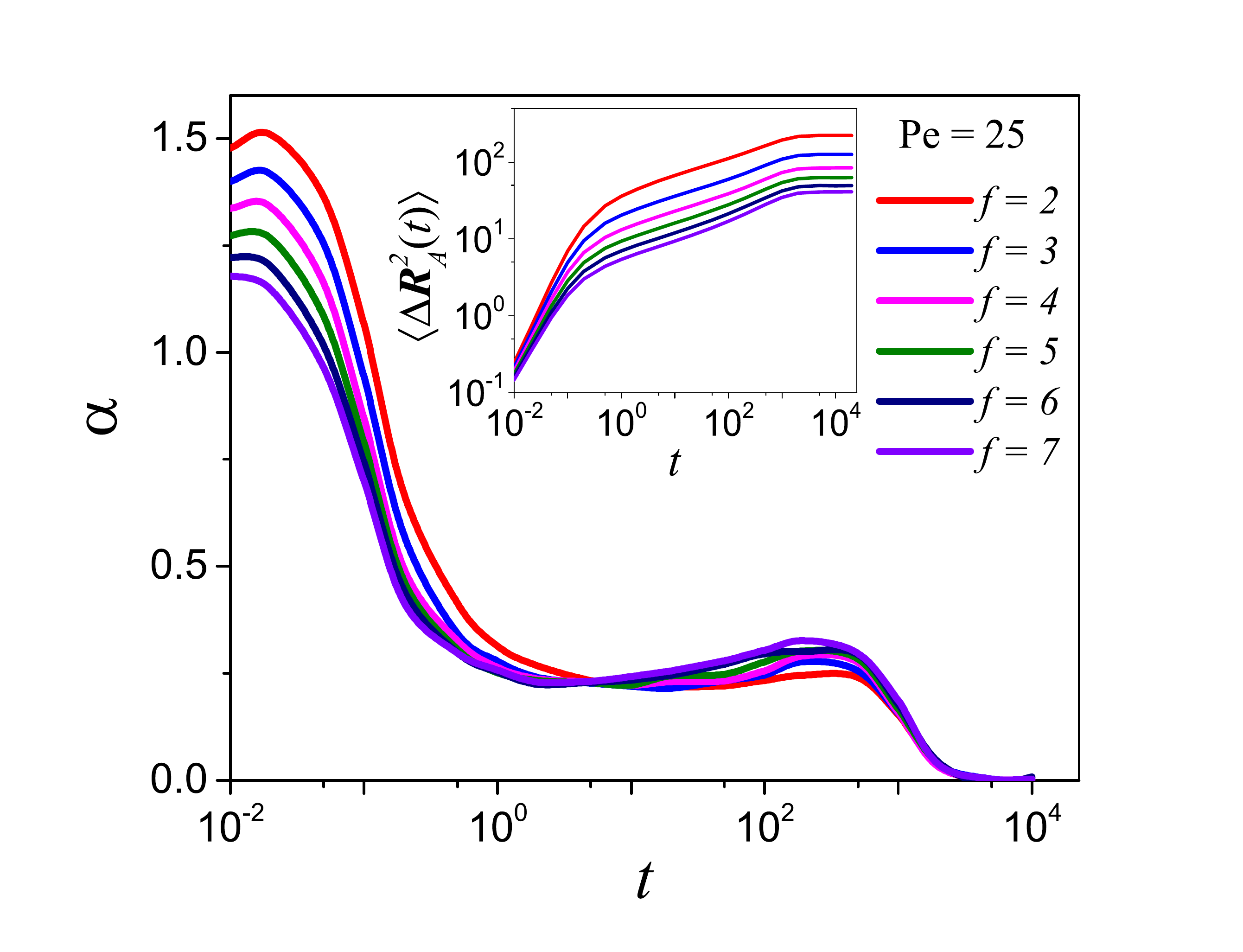}
	\caption{ The variation of the anomaly exponent $\alpha(t)=d\log\mathrm{MSD}/d\log t$ for the ABP cross-linker for varying the functionality $f$ from 2 to 7. The inset shows the corresponding MSD curves. In the simulation, the P{\'e}clet number is $\mathrm{Pe} = 25$.}
	\label{fgr:msd-vf}
\end{figure}

For the Brownian cross-linker, it was shown by an analytical study that by changing $f$ the MSD grows as $\langle \Delta\mathbf{R}^2(t)\rangle\sim  \frac{k_BT}{f}t^{1/2}$ at the intermediate time scale and reaches a plateau $\frac{6k_BT N}{fk}$ at the stationary state \cite{sprakelPRL}. This demonstrates that at $\mathrm{Pe}=0$ the anomaly exponent is invariant to $f$, but the amplitude of MSD is reduced by a factor $f$. To investigate the case for the ABP cross-linker, we simulate our model by changing $f$ from 2 to 7.  Fig.~\ref{fgr:msd-vf} presents the variation of $\alpha(t)=d\log \mathrm{MSD}(t)/d\log t$ for the active cross-linkers connected with the network having $f=2,~3,\ldots,7$ at a fixed $\mathrm{Pe}$. The corresponding MSD curves are plotted in the inset. The simulation results suggest that, analogous to the BP limit, the functionality does not change the overall feature of the ABP dynamics in the Rouse and long-time regimes; after the initial superdiffusion region the MSDs keep the same scaling relation of $\sim t^\alpha$ with $\alpha\approx 0.23$, regardless of $f$, for more than two decades. The MSDs (inset) tell that the functionality has a role of decreasing the amplitude of MSD, particularly, its stationary value.

The effect of $f$ on the stationary state can be understood by the analogy with a single ABP confined to a harmonic potential of stiffness $K_{\mathrm{eff}}\propto f/N$. In this coarse-grained picture, the motion of an ABP trapped in a harmonic potential $U(x)=\frac{1}{2}K_{\mathrm{eff}} x^2$ is described by the Langevin equation
\begin{equation}\label{eq:plateau}
	\gamma_\mathrm{eff}\dot{\mathbf{R}}_A=-K_{\mathrm{eff}}\mathbf{R}_A(t)+\mathbold{\xi}(t)+\mathbold{\eta}'(t).
\end{equation}
Here, via the interactions with the polymer system, the particle may have an effective frictional coefficient $ \gamma_\mathrm{eff}$ and an active noise characterized by $\langle \eta'_\mu(t)\eta'_{\nu}(t')\rangle=\frac{\gamma_\mathrm{eff}^2  v_\mathrm{p}^2 }{3}\delta_{\mu,\nu}\exp{(-|t-t'|/\tau_{\mathrm{eff}})} $. We solve this equation~\cite{um2019},  identifying that  \begin{equation}\label{eq:MSDst_SM}
	\lim_{t\rightarrow\infty}\langle\Delta \mathbf{R}_A^2\rangle\equiv \langle\Delta \mathbf{R}_A^2\rangle_{\mathrm{st}}= 3\frac{k_BT}{K_\mathrm{eff}}+\frac{ v_\mathrm{p}^2}{\frac{K_\mathrm{eff}}{\gamma_\mathrm{eff}}\left(\frac{K_\mathrm{eff}}{\gamma_\mathrm{eff}}+\frac{1}{\tau_{\mathrm{eff}}}\right)}.
\end{equation} In accordance with this theory, our simulation easily confirms that $\langle\Delta \mathbf{R}_A^2(\mathrm{Pe}=0)\rangle_{\mathrm{st}}\sim f^{-1}$ (Fig.~S3) with $K_{\mathrm{eff}}=\frac{fk}{2N}$. Intriguingly, the second term signifies that its $f$-dependence is multi-scaling, depending on the ratio of $1/\tau_\mathrm{eff}$ to $K_\mathrm{eff}/\gamma_\mathrm{eff}$; if $\tau_\mathrm{eff}\gg \gamma_\mathrm{eff}/K_\mathrm{eff}$, then the second term scales as $1/f^2$, whereas for the opposite limit it scales as $1/f$. It is determined that for given simulation conditions the data approximately follow $\sim f^{-2}$ where the fit gives $K_\mathrm{eff}/\gamma_\mathrm{eff}=1.97f$ and $\tau_\mathrm{eff}=0.782$ (Fig.~S4).

\subsection{The effect of boundary condition}
\label{subsec:boundary}
\begin{figure}
	\centering
	\includegraphics[width=8.5cm]{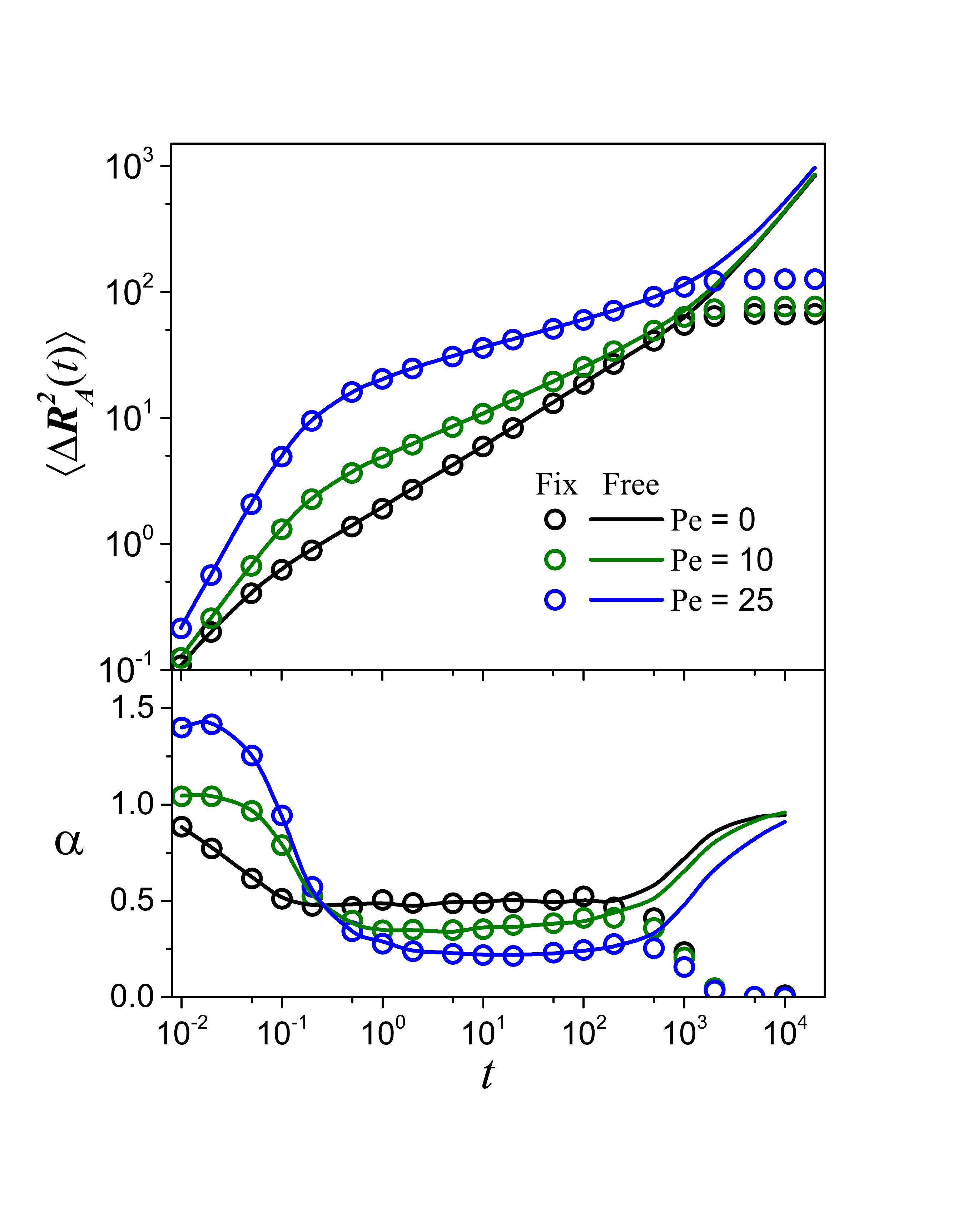}
	\vspace{-1.5 cm}
	\caption{The comparison of the dynamics of the ABP cross-linkers for fixed and free boundary conditions. (Top) The MSD curves. (Bottom) The anomaly exponents $\alpha$ as a function of $t$. In both panels, the symbols and solid lines represent the simulation data for the fixed and free boundary conditions, respectively. In the plot, the functionality is $f=3$ and $\mathrm{Pe =0}$ (black), $10$ (green), and $25$ (blue).  }
	\label{fig:MSD_Fix_Free}
\end{figure}

We repeat the previous simulations with the free boundary conditions, where the end beads of the chains are moving ($\frac{\partial \mathbf{R}_N}{\partial N}=0$). In Fig.~\ref{fig:MSD_Fix_Free} we compare the corresponding MSDs and $\alpha(t)$ of this model with the above results for the fixed boundary conditions. The plot shows that the two ABPs display exactly identical dynamics, independently of the boundary conditions, up to the Rouse regime. The only difference is observed in the long-time regime: here, unlike the fixed boundary condition, the free case allows the ABP to be Brownian at large times. The plotted three MSDs at $\mathrm{Pe}=0,~ 10$, and $25$ approach one another in this regime, signifying that the long-time Fickian dynamics is attributed to the drift of the entire system itself. 
A supplementary simulation study shows that the ABP cross-linker has the boundary-free anomalous dynamics until the length of the arm is as short as $N\sim O(10)$ [see Fig.~S5 for further discussion].  
It is informative to learn that the boundary condition gives trivial effects on the ABP dynamics. The observed dynamic properties for the ABP cross-linker above, such as the gaussianity, autocorrelation structure, and the cross-correlation with the beads in the polymer chain, will be preserved independently of the boundary condition. 

\section{Results: analytic theory}
\label{sec:theory}

Based on the simulation studies in the previous section, we now analytically revisit the athermal dynamics of the ABP-polymer systems, starting from the Langevin equation \eqref{eq:method_abp}, and establish a quantitative theory on the anomalous dynamics for the ABP and the passive particles in the polymer. For analytic tractability, hereafter we consider a minimal model where the active cross-linker is connected with two arms ($f=2$) having free ends. Note that the study of this minimal system is sufficient for understanding the anomalous dynamics for the systems observed in the simulation above, since, as demonstrated in Sec.~\ref{subsec:f} and \ref{subsec:boundary}, the number of $f$ and the boundary conditions are irrelevant to this issue. To proceed with the calculation, we rewrite the set of the original Langevin equation~\eqref{eq:method_abp} into the following unified form:
\begin{equation}\label{eq:theory_abp}
\gamma\frac{d\mathbf{R}_{s}}{dt}=-k(2\mathbf{R}_{s}-\mathbf{R}_{s+1}-\mathbf{R}_{s-1})+\mathbold{\xi}_{s}(t)+\mathbold{\eta}_s(t) 
\end{equation}
where $\mathbf{R}_0(=\mathbf{R}_A)$ and $\mathbf{R}_{s\neq0}$ represent the ABP cross-linker and the other Brownian beads in the polymer, respectively, with the bead index $s$ running from $-N$ to $N$.  In the above equation the last term $\mathbold{\eta}_s(t)$  is responsible for the active noise acting on the bead at $s$. In our study, it is specified to be $\mathbold{\eta}_s(t)=\delta_{s,0}\mathbold{\eta}(t)$ so that the active noise is only applied to the center particle. Through the modification of this selectivity index, the above model also describes other active polymer systems, e.g., where multiple beads in the polymer are the ABPs~\cite{osmanovic2018_JCP} or all beads are active \cite{osmanovic2017,Samanta2016_JPA}. For a better understanding of our current model, we have also solved the latter model, such that all particles are the ABPs. It turns out that the latter system is trivial in that the cross-linker always exhibits Rouse-like dynamics $\langle \Delta \mathbf{R}^2\rangle \sim t^{1/2}$ for any P{\'e}clet numbers [Fig.~S2]. This model is discussed in Sec.~II B in the ESI\dag. 

Now, exploiting the technique of normal mode expansion~\cite{osmanovic2017,sakaue2017_actpol,osmanovic2018_JCP,gov2014_BPJ}
\begin{equation}\label{eq:cos_transform}
\begin{aligned}
&\mathbf{R}_{s}=\tilde{\mathbf{X}}_0+2 \sum^{2N}_{p=1} \tilde{\mathbf{X}}_{p} \cos{\left(\frac{p \pi (s+N)}{2N}\right)}\\ 
&\tilde{\mathbf{X}}_{p}=\frac{1}{2N} \sum^{N}_{s=-N} \mathbf{R}_{s} \cos{\left(\frac{p \pi (s+N)}{2N}\right)}
\end{aligned}
\end{equation}
we recast Eq.~\eqref{eq:theory_abp} into the Langevin equation of each modes, as in the following:
\begin{equation}\label{eq:Langevin_TF}
\gamma\frac{d\tilde{\mathbf{X}}_{p}(t)}{dt} = k_{p} \tilde{\mathbf{X}}_{p}(t) + \tilde{\mathbold{\xi }}_{p}  (t) + \tilde{\mathbold{\eta}}_{p} (t).
\end{equation}
Here, the spring constant becomes $k_p=4k\sin^2{\left(\pi p /4N \right)} \approx k \pi^2 p^2 /4 N^2$ in the large-$N$ limit while keeping $k/N^2$ finite, and $\tilde{\mathbold{\xi}}_p$ and $\tilde{\mathbold{\eta}}_p$ are, respectively, the cosine-transformed noises of the original noises. The transformed thermal noise has the autocorrelation 
\begin{equation}\label{eq:Thauto_TF}
\langle \tilde{\mathbold{\xi}}_{p} (t) \cdot \tilde{\mathbold{\xi}}_{q} (t')\rangle = \frac{3k_BT\gamma}{2N}\delta_{p,q}(1+\delta_{p,0})\delta(t-t'),
\end{equation}
while the active noise has an exponentially decaying, mode-coupled autocorrelation 
\begin{equation}\label{eq:OUauto_TF}
\langle \tilde{\mathbold{\eta}}_p(t)\cdot\tilde{\mathbold{\eta}}_{q}(t')\rangle =\frac{\gamma^2  v_\mathrm{p}^2}{4N^2} \exp{(-|t-t'|/\tau_A)} \cos{\left(\frac{p \pi}{2}\right)}\cos{\left(\frac{q \pi}{2}\right)}.
\end{equation}
The stationary solution of the Langevin equation \eqref{eq:Langevin_TF} is simply given by
\begin{equation}\label{eq:Langevin_solution}
\tilde{\mathbf{X}}_{p}(t) = \frac{1}{\gamma}\int^{t}_{-\infty}dt' e^{-\frac{(t-t')p^2}{\tau_R}} \left( \tilde{\mathbold{\xi}}_p (t')+\tilde{\mathbold{\eta}}_p (t') \right),
\end{equation}
where the integral starts from negative infinity in order to eliminate the initial condition effect and where $\tau_R=4\gamma N^2/(\pi^2 k)$ is the Rouse relaxation time. From Eqs.~\eqref{eq:Thauto_TF} and \eqref{eq:OUauto_TF}, we find that $\tilde{\mathbf{X}}_{p}$ has the autocorrelation ($t>t'$)
\begin{equation}\label{eq:Xp_autocorrelation}
\begin{aligned}
\langle\tilde{\mathbf{X}}_{p\neq0}(t) \cdot \tilde{\mathbf{X}}_{q\neq0}(t') \rangle &= \frac{3 k_BT}{2N\gamma} \delta_{p,q} \frac{\tau_R{\rm e}^{-\frac{p^2}{\tau_R}(t-t')}}{2p^2}  \\ 
&+\frac{v_\mathrm{p}^2\tau_R}{4N^2}\frac{\cos\left(\frac{p\pi}{2}\right)\cos\left(\frac{q\pi}{2}\right)}{p^2+q^2}\left[\frac{{\rm e}^{-\frac{p^2}{\tau_R}(t-t')}}{\frac{p^2}{\tau_R}+\frac{1}{\tau_A}}\right. \quad \\
&\quad+\left.
\frac{{\rm e}^{-\frac{t-t'}{\tau_A}}}{\frac{q^2}{\tau_R}+\frac{1}{\tau_A}}+\frac{{\rm e}^{-\frac{t-t'}{\tau_A}}-{\rm e}^{-\frac{p^2}{\tau_R}(t-t')}}{\frac{p^2}{\tau_R}-\frac{1}{\tau_A}}\right]
\end{aligned}
\end{equation}
and
\begin{equation}
\langle\tilde{\mathbf{X}}_{0}(t) \cdot \tilde{\mathbf{X}}_{0}(t') \rangle = - \frac{v_\mathrm{p}^2\tau_A^2}{4N^2}{\rm e}^{-\frac{t-t'}{\tau_A}}+\left(\frac{3 k_BT}{N\gamma} + \frac{v_\mathrm{p}^2\tau_A}{2N^2}\right) \int_{-\infty}^{t'} {\rm d}u.
\end{equation}
In this expression, the first line in the R.H.S. describes the autocorrelation for the usual passive Rouse chain. Then the remaining part solely explains the ABP-driven effect, which is inclusive of a coupling between the modes and vanishes for all odd numbers of $p ~\hbox{and}~ q \in\{\pm 1, \pm 3, \pm 5, \ldots\}$. 

\subsection{General expression for MSD}
\begin{figure}
	\centering
	\includegraphics[width=8.5cm]{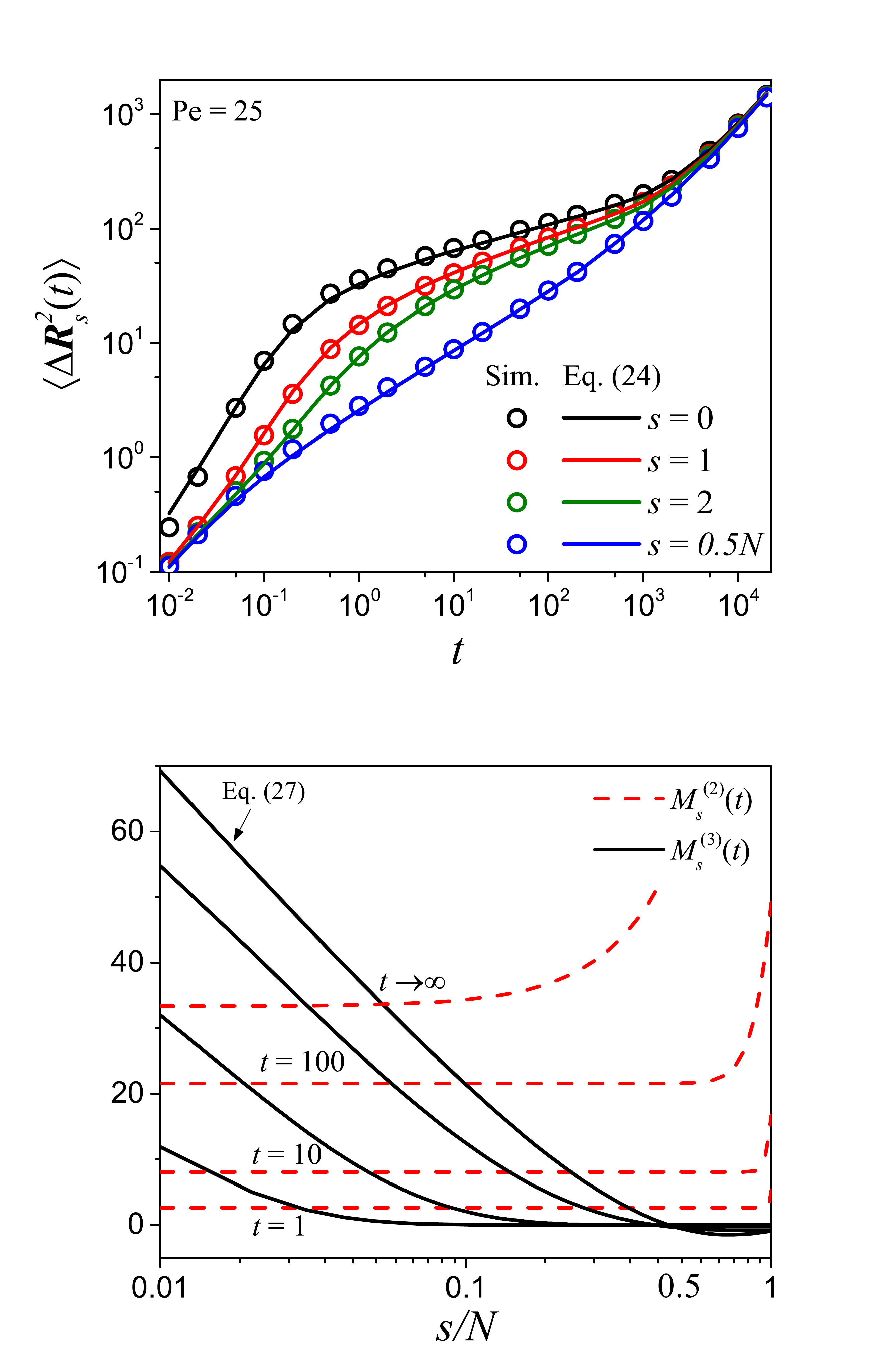}
	\caption{ (Top) Comparison between the theoretical prediction from Eq.~\eqref{eq:MSD2} and the simulation results for MSD. The plotted MSDs are for the active cross-linker ($s=0$) and the Brownian beads in the arms at $s=1$, 2, and $0.5N$ (where $N=100$). The theoretical lines (solid) are obtained by numerically evaluating Eq.~\eqref{eq:MSD2}. In the simulation, we use $f=2$ and $\mathrm{Pe}=25$. (Bottom) The plots of $M^{(3)}_s(t)$ [Eq.~\eqref{eq:MSD-3-redux}] and $M^{(2)}_s(t)$ [Eq.~\eqref{eq:Ms2}] as a function of $s$ for $t=1,~10,~100$, and the limit $t\to\infty$.  }
	\label{fig8}
\end{figure}

Within the normal mode expansion the MSD of a bead in the ABP-polymer system is formally expressed as 
\begin{equation}\label{eq:MSD}
\langle \Delta\mathbf{R}^2_s (t)\rangle 
=\left\langle \left[\Delta\tilde{\mathbf{X}}_0 (t)  + 2 \sum^{\infty}_{p=1} \Delta\tilde{\mathbf{X}}_{p}(t)  \cos\left(\frac{ p \pi (s+N)}{2N} \right)\right]^2\right\rangle
\end{equation}
where $\Delta\mathbf{R}_s(t)=\mathbf{R}_s (t) - \mathbf{R}_s(0)$, $\Delta\tilde{\mathbf{X}}_p (t)=\tilde{\mathbf{X}}_p (t) - \tilde{\mathbf{X}}_p (0) $, and the number of modes are taken to be infinite in the large-$N$ limit imposed below Eq.~\eqref{eq:Langevin_solution}. Making use of the autocorrelation relations Eq.~\eqref{eq:Xp_autocorrelation}, we can obtain the analytic expression for MSD by the following steps: (i) The square average of $\Delta\tilde{\mathbf{X}}_0 (t)$ is responsible for the drift of the entire system, which is easily evaluated to 
\begin{equation}\label{eq:MSD-1}
\begin{aligned}
M_s^{(1)}(t)\equiv \left\langle \Delta\tilde{\mathbf{X}}^2_0 (t)\right\rangle 
=\frac{3 D}{N} t + \frac{v^2_\mathrm{p} \tau_A}{2 N^2} \left(\tau_A \left(e^{-t/\tau_A}-1\right) + t\right).
\end{aligned}
\end{equation}
Here, the first and second terms explain the thermal ($D=k_BT/\gamma$) and ABP-driven drift, respectively. For $t\gg \tau_A$, both terms grows linearly with $t$, as expected. (ii) The square average of the cross product reads  \begin{equation}\label{eq:MSD-2}
\begin{aligned}
M_s^{(a)}(t) &\equiv4 \sum^{\infty}_{p=1} \left<\Delta\tilde{\mathbf{X}}_0 (t) \cdot \Delta\tilde{\mathbf{X}}_{p}(t) \right> \cos\left(\frac{ p \pi (s+N)}{2N} \right)\\
&=\sum^{2N}_{p=1} \frac{\tau_R v_\mathrm{p}^2}{N^2 p^2} \cos{\left(\frac{p \pi}{2}\right)} \cos{\left(\frac{p \pi (s+N)}{2N}\right)} \\
&~~\quad \times\left[\frac{2-e^{-\frac{p^2 t}{\tau_R}}-e^{-\frac{t}{\tau_A}}}{\frac{1}{\tau_A}+\frac{p^2}{\tau_R}} \right.  \left. - \frac{e^{-\frac{p^2 t}{\tau_R}} -e^{-\frac{t}{\tau_A}}}{\frac{1}{\tau_A}-\frac{p^2}{\tau_R}}\right]. 
\end{aligned}
\end{equation}
(iii) The square average of the last term is
\begin{equation}
\begin{aligned}\label{eq:MSD-3}
M_s^{(b)}(t)&\equiv 4\sum^{\infty}_{p=1} \sum^{\infty}_{q=1} \left<\Delta\tilde{\mathbf{X}}_p (t) \cdot \Delta\tilde{\mathbf{X}}_{q}(t) \right>\\& \times\cos\left(\frac{ p \pi (s+N)}{2N} \right)\cos\left(\frac{ q \pi (s+N)}{2N} \right)\\
&= M_s^{(2)}(t)+ M_s^{(c)}(t)
\end{aligned}
\end{equation}
where 
\begin{equation}\label{eq:Ms2}
M_s^{(2)}(t)= \sum^{\infty}_{p=1}\frac{6 D \tau_R}{N  p^2}\left(1-e^{-\frac{p^2 t}{\tau_R}}\right)\cos^2{\left(\frac{p\pi (s+N)}{2N}\right)}
\end{equation}
and
\begin{equation}
\begin{aligned}
M_s^{(c)}(t)&=\sum^{\infty}_{p=1}\sum^{\infty}_{q=1}\frac{2\tau_R v_\mathrm{p}^2}{N^2(p^2+q^2)} \cos{\left(\frac{p\pi}{2}\right)}\cos{\left(\frac{q\pi}{2}\right)} \\ &\times\cos{\left(\frac{p\pi (s+N)}{2N}\right)}\cos{\left(\frac{q\pi (s+N)}{2N}\right)}\\
&\times\left[\frac{1-e^{-\frac{p^2 t}{\tau_R}}}{\frac{1}{\tau_A}+\frac{p^2}{\tau_R}} + \frac{1-e^{-\frac{t}{\tau_A}}}{\frac{1}{\tau_A}+\frac{q^2}{\tau_R}} + \frac{e^{-\frac{t}{\tau_A}}-e^{-\frac{p^2 t}{\tau_R}}}{\frac{1}{\tau_A} - \frac{p^2}{\tau_R}}\right].
\end{aligned}
\end{equation}
Here, $M_s^{(c)}(t)$ can be further simplified after performing a summation over $q$ using the Euler-Maclaurin formula $\sum_{k=1}^\infty f(k)\approx  \int_0^\infty f(k){\rm d}k -f(0)/2$. Then $M_s^{(a)}(t)$ and $M_s^{(c)}(t)$ can be summed up together to give
\begin{equation}
\begin{aligned}\label{eq:MSD-3-redux}
M_s^{(3)}(t)=&M_s^{(a)}(t)+M_s^{(c)}(t) \\
=&\sum^{\infty}_{p=1}\frac{\pi\tau_R v_\mathrm{p}^2}{4pN^2} \cos{\left(\frac{p\pi s}{N}\right)}\exp\left(-\frac{p\pi s}{N}\right)\\&\times \left[\frac{2-e^{-\frac{4p^2 t}{\tau_R}}-e^{-\frac{t}{\tau_A}}}{\frac{1}{\tau_A}+\frac{4p^2}{\tau_R}} + \frac{e^{-\frac{t}{\tau_A}}-e^{-\frac{4p^2 t}{\tau_R}}}{\frac{1}{\tau_A} - \frac{4p^2}{\tau_R}}\right].
\end{aligned}
\end{equation}
Collecting the three terms, we now obtain one of the main results in this section, i.e. the MSD of an arbitrary bead in the polymer, as
\begin{equation}\label{eq:MSD2}
\langle \Delta\mathbf{R}^2_s (t)\rangle 
=M_s^{(1)}(t)+M_s^{(2)}(t)+M_s^{(3)}(t).
\end{equation}
We perform the Langevin simulations for the corresponding system and numerically validate Eq.~\eqref{eq:MSD2} with them. Fig.~\ref{fig8} shows the comparison between Eq.~\eqref{eq:MSD2} and the simulation data (open circles). Here, the solid lines depict Eq.~\eqref{eq:MSD2} numerically estimated for the beads at $s=0,~1,~2$, and $N/2$. Excellent agreement between them is observed over the entire time window plotted.

The above general expression for MSD provides the following relevant information about the polymer dynamics: First, in the limit of $v_\mathrm{p}\to0$ ($\mathrm{Pe}=0$) our ABP-polymer system becomes a Rouse chain, at which the MSD \eqref{eq:MSD2} is reduced to   
\begin{equation}\label{eq:MSD-Rouse}
\begin{aligned}
\langle \Delta\mathbf{R}^2_s (t)\rangle &=
\frac{3D}{N}t + \sum^{\infty}_{p=1}\frac{6D\tau_R}{N  p^2}\left(1-e^{-\frac{p^2 t}{\tau_R}}\right)\cos^2\left(\frac{p\pi (s+N)}{2N}\right) \\
&\equiv M_\mathrm{Rouse}(t).
\end{aligned}
\end{equation}
This is exactly the expression of the MSD for a bead in the Rouse chain~\cite{rubinstein2003polymer}. In the above expression, we can evaluate the summation over $p$ by using again the Euler-Maclaurin formula. As an example, the MSD for the center bead is easily calculated to be
\begin{equation}\label{eq:MSD-vp0}
\begin{aligned}
\langle \Delta\mathbf{R}^2_0(t)\rangle=\frac{3D}{N}t+\frac{3D}{N}\sqrt{\pi\tau_Rt}\left(1-\text{erf}\left(\sqrt{\frac{4t}{\tau_R}}\right)\right)\\+\frac{9D\tau_R}{4N}\left(1-e^{-\frac{4t}{\tau_R}}\right) .
\end{aligned}
\end{equation}
It can be shown that for $t\ll \tau_R$ the MSD exhibits the Rouse dynamics where $\langle \Delta\mathbf{R}^2_0(t)\rangle\simeq \sqrt{\frac{9\pi \tau_R D^2}{N^2}}t^{1/2}$ and for $t\gg \tau_R$ the drift dominates over the Rouse dynamics, where $\langle \Delta\mathbf{R}^2_0(t)\rangle\simeq\frac{3D}{N}t$. 

Second, for $\mathrm{Pe}\neq0$, the effect of the active noise is incorporated in $M_s^{(1)}(t)$ [Eq.~\eqref{eq:MSD-1}] and $M_s^{(3)}(t)$ [Eq.~\eqref{eq:MSD-3-redux}]. As the former part describes the drift of the entire system, only the latter is responsible for the polymer internal dynamics driven by the ABP. To understand these ABP-driven internal dynamics, in Fig.~\ref{fig8} (Bottom) we numerically plot $M_s^{(3)}(t)$ as a function of $s$ at several fixed time $t$. As seen in this plot, the effect of the central active noise rapidly decays out with $s$ for all $t$. For $t\gg \tau_R$, $M_s^{(3)}(t)$ approaches towards the limiting curve 
\begin{equation}\label{eq:MSD-3-tinfty}
\begin{aligned}
\lim_{t\rightarrow\infty}M_s^{(3)}(t)=\frac{\pi\tau_R \tau_Av_\mathrm{p}^2}{2N^2}\left(\frac{\pi s}{2N}-{\rm ln}\sqrt{2{\rm cosh}\left(\frac{\pi s}{N}\right)-2{\rm cos}\left(\frac{\pi s}{N}\right)}\right)
\end{aligned}
\end{equation} 
obtained at the condition of $\tau_A\ll\tau_R$ and $s>0$ (see this curve in the plot). From this limiting curve we can see that $M_s^{(3)}(t)$ decreases with $s$ as $\frac{\pi\tau_R \tau_Av_\mathrm{p}^2}{2N^2}\left[\frac{\pi s}{2N}-{\rm ln}\left(\frac{\sqrt{2}\pi s}{N}\right)\right]$ from the center. Note that this logarithmic decay is faster than a linear decay in the range of the plotted $s/N$. We also remark that $M_s^{(3)}(t)$ goes to zero at a certain value of $s=s_\mathrm{zero}(t)$, whose value is always less than $s=N/2$ for all $t$ [note: the maximal zero-crossing value at $t\to\infty$ reads $s_\mathrm{zero}/N\approx0.4627$ from Eq.~\eqref{eq:MSD-3-tinfty}]. After the zero-crossing point $s_\mathrm{zero}$, $M_s^{(3)}$ has a negative dip. But, the magnitude of this dip is smaller than 10\% of the initial maximum value of  $M_s^{(3)}$. Therefore, it can be concluded that, roughly speaking, the effect of the active perturbation on the bead $s=0$ can be neglected for $s\gtrsim N/2$. For our simulation in the previous section, this is the main reason why the bead dynamics rapidly follows the Rouse dynamics $M_\mathrm{Rouse}(t)$ as $s\to N/2$ in Fig.~\ref{fgr:monomer} (as well as in Fig.~\ref{fig8} (Top)).

In Fig.~\ref{fig8} (Bottom), we plot $M_s^{(2)}$ (red dashed lines) for several $t$ [here the sharp increase at near $s=N$ is from the boundary effect, otherwise $M_s^{(2)}$ is independent of $s$, see Eq.~\eqref{eq:MSD-vp0} and below]. The plot shows that for given $t$ there is a crossing of the two curves $M_s^{(3)}$ and $M_s^{(2)}$. Hence, we are allowed to define a characteristic distance $s^*$ equating $M_{s^*}^{(3)}= M_{s^*}^{(2)}$; for $s>s^*$ the beads are dominated by the Rouse (thermal) dynamics while, for the opposite condition, highly affected by the ABP-driven dynamics $M_s^{(3)}$. For given $v_\mathrm{p}$ and with a large-time limit $t\gg \tau_R$, this characteristic distance is obtained by this relation 
\begin{equation}
v_\mathrm{p}^{2} = \frac{\pi ND}{4\tau_A}  \left|\frac{\pi s^*}{2N}-{\rm ln}\sqrt{2{\rm cosh}\left(\frac{\pi s^*}{N}\right)-2{\rm cos}\left(\frac{\pi s^*}{N}\right)}\right|^{-1}.
\end{equation}
We can see from this figure that $s^*(t,v_\mathrm{p})$ is much smaller than $N/2$ at any time $t$. This means that the polymer beads quickly approach the Rouse dynamics even before $s=N/2$, as is consistent with the simulation results in Fig.~\ref{fgr:monomer}.

\subsection{Dynamics of the ABP cross-linker}\label{sec:ABP}
\begin{figure}
	\centering
	\includegraphics[width=8.5cm]{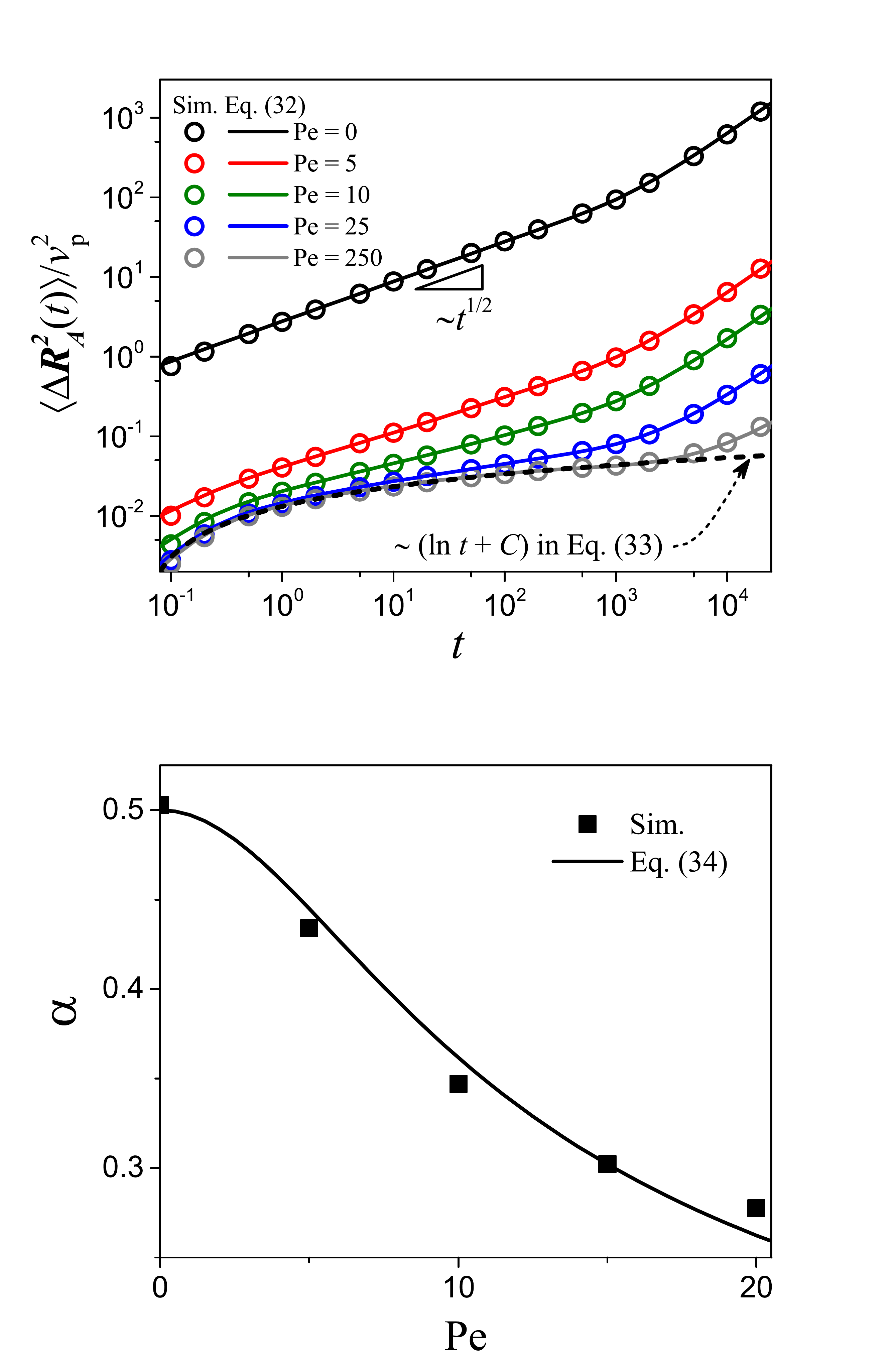}
	\caption{Anomalous diffusion for ABP cross-linkers in the polymer with $f=2$, $N=100$, and with the free ends. (Top) MSD curves for $\mathrm{Pe}=0, ~5, ~10,~25$, and $250$. The solid lines depict the theoretical prediction from  Eq.~\eqref{eq:MSDABP}, for the given parameter values. The dashed line depicts the prediction from Eq.~\eqref{eq:MSDSim}. (Bottom) The apparent anomaly exponent $\alpha$ in the Rouse regime as a function of $\mathrm{Pe}$. The data (filled square) are extracted from the MSDs (Top) by a power-law fit $\sim t^\alpha$ within $[1, 500]$. The solid line represents Eq.~\eqref{eq:alpha}. See the text for further information. }
	\label{fig9}
\end{figure}

Now let us focus on the dynamics of the ABP cross-linker and find the compact analytic form of its MSD. Formally, the corresponding MSD is obtained from Eq.~\eqref{eq:MSD2} by setting $s=0$. Since $M_s^{(1)}(t)$ and $M_s^{(2)}(t)$ are already obtained, our task is to obtain the expression for $M_{0}^{(3)}(t)$. Below we briefly explain the procedure. Inserting $s=0$ into Eq.~\eqref{eq:MSD-3-redux} and rearranging the terms in the parenthesis, we can recast $M_0^{(3)}(t)$ into the following expression:
\begin{equation}
\begin{aligned}\label{eq:MSD-4}
M_0^{(3)}(t)=&\sum^{\infty}_{p=1}\frac{\pi\tau_R\tau_A v_\mathrm{p}^2}{2N^2} \left[\frac{1-e^{-\frac{4p^2 t}{\tau_R}}}{p}-\frac{2p}{\tau_R}\frac{2}{\frac{1}{\tau_A}+\frac{p^2}{\tau_R}} \right.\\
&~~ + \left.\frac{2p}{\tau_R}\left(\frac{e^{-\frac{4p^2 t}{\tau_R}}+e^{-\frac{t}{\tau_A}}}{\frac{1}{\tau_A}+\frac{4p^2}{\tau_R}} + \frac{e^{-\frac{t}{\tau_A}}-e^{-\frac{4p^2 t}{\tau_R}}}{\frac{1}{\tau_A} - \frac{4p^2}{\tau_R}}\right)\right].
\end{aligned}
\end{equation}
Then the summation is carried out using the Euler-Maclaurin formula. From this procedure, the first line of Eq.~\eqref{eq:MSD-4} is evaluated to
\begin{equation}
\label{eq:MSD-4a}
\frac{\pi\tau_R\tau_A v_\mathrm{p}^2}{4N^2}\left[-\Gamma\left(0,\frac{4t}{\tau_R}\right)+1-{\rm e}^{-\frac{4t}{\tau_R}}+{\rm ln}\left(\frac{\tau_R}{4\tau_A}+1\right)-\frac{4\tau_A}{\tau_R+4\tau_A}\right]
\end{equation}
with the incomplete gamma function $\Gamma(z,x)=\int_x^\infty  t^{z-1}e^{-t} dt$. For the second line of Eq.~\eqref{eq:MSD-4}, we perform the integration for both terms simultaneously with the condition of  $\tau_A<4\tau_R$ and find the result
\begin{equation} 
\label{eq:MSD-4b}
\begin{aligned}
&\frac{\pi\tau_R\tau_A v_\mathrm{p}^2}{8N^2}\left[{\rm e}^{t/\tau_A}\Gamma\left(0,\left(\frac{1}{\tau_A}+\frac{4}{\tau_R}\right)t\right)\right.\\
&~~ \left.-{\rm e}^{-t/\tau_A}{\rm Ei}\left(\left(\frac{1}{\tau_A}-\frac{4}{\tau_R}\right)t\right)+{\rm ln}\left(\frac{\frac{1}{\tau_A}-\frac{4}{\tau_R}}{\frac{1}{\tau_A}+\frac{4}{\tau_R}}\right)\right].
\end{aligned}
\end{equation}
Here, $\mathrm{Ei}(x)=-\int_{-x}^\infty dt e^{-t}/t $ is the exponential integral. In the limit of $t\gg\tau_A$ this term decays to a constant, which further becomes negligible if $\tau_R \gg 4\tau_A$. Collecting the expressions from Eqs.~\eqref{eq:MSD-1}, \eqref{eq:MSD-vp0}, \eqref{eq:MSD-4a}, and \eqref{eq:MSD-4b}, we obtain the compact expression for the MSD of the cross-linker as below: 
\begin{equation}\label{eq:MSDABP}
\begin{aligned}
&\langle \Delta\mathbf{R}^2_A (t)\rangle = \frac{3 D}{N} t + \frac{v^2_\mathrm{p} \tau_A}{2 N^2} \left(\tau_A \left({\rm e}^{-t/\tau_A}-1
\right) + t\right)\\
&+\frac{3D}{2N}\left(2\sqrt{\pi \tau_R t} \left(1-\textrm{erf}\left(\sqrt{\frac{4t}{\tau_R}}\right)\right) - \frac{3\tau_R}{2} \left( {\rm e}^{-\frac{4t}{\tau_R}}-1\right)\right) \\
&+\frac{\pi\tau_R\tau_A v_\mathrm{p}^2}{4N^2}\left[-\Gamma\left(0,\frac{4t}{\tau_R}\right)-{\rm e}^{-\frac{4t}{\tau_R}}+\frac{1}{2}{\rm ln}\left(\frac{\tau_R^2}{16\tau_A^2}-1\right)+\frac{\tau_R}{\tau_R+4\tau_A}\right.\\
&\qquad + \left.\frac{{\rm e}^{t/\tau_A}}{2}\Gamma\left(0,\left(\frac{1}{\tau_A}+\frac{4}{\tau_R}\right)t\right) -\frac{{\rm e}^{-t/\tau_A}}{2}{\rm Ei}\left(\left(\frac{1}{\tau_A}-\frac{4}{\tau_R}\right)t\right)\right].
\end{aligned}
\end{equation}
We numerically validate Eq.~\eqref{eq:MSDABP} with the corresponding simulation. For this purpose, in Fig.~\ref{fig9}, we plot the rescaled MSDs $\langle \Delta\mathbf{R}^2_A (t)\rangle/v_\mathrm{p}^2$ of the ABP cross-linker at various P{\'e}clet numbers. Here, the solid curves are the predictions from Eq.~\eqref{eq:MSDABP} with the given parameter values, which show excellent agreement with the data for all tested $\mathrm{Pe}$.

Based on Eq.~\eqref{eq:MSDABP}, let us look for the origin of the anomalous diffusion $\sim t^{\alpha}$ for the ABP motion observed in the Rouse regime in the simulation. To find a clue we expand the above expression at the timescale $\tau_A\ll t\ll \tau_R$, which yields
\begin{equation}\label{eq:MSDSim}
\begin{aligned}
\langle \Delta&\mathbf{R}^2_A (t)\rangle \approx \frac{\tau_A\tau_R v_\mathrm{p}^2 \pi}{4N^2} \left(A t + B \sqrt{t} + {\rm ln}~ t  +C \right).
\end{aligned}
\end{equation}
In the above, the expansion coefficients read $A=\frac{2+8\pi}{\pi\tau_R}$, $B=\frac{12DN}{\tau_Av_\mathrm{p}^2\sqrt{\pi\tau_R}}$, and $C=0.5772+{\rm ln}\left(\frac{1}{\tau_A}+\frac{4}{\tau_R}\right)-\frac{4\tau_A}{\tau_R+4\tau_A}-\frac{2\tau_A}{\pi\tau_R}$. 
From this result we can observe the two important findings for the ABP dynamics: (1) The anomalous diffusion of the active cross-linker is actually not simply governed by a single power-law increment ($\sim t^\alpha$). It fact, it is apparent dynamics, only phenomenologically valid over a certain time window, originating from the superimposed motion of the drift ($\sim t$), Rouse ($\sim t^{1/2}$), and the remaining collective ($\sim \ln t$) dynamics. 
(2) The ABP interacting with a passive polymer network attains logarithmic dynamics in its MSD through the viscoelastic feedback effect. This is a very interesting finding; if the polymer was made up of identical ABPs, as shown in some work (including our calculation)~\cite{osmanovic2017}, the corresponding term would change from $\mathrm{ln}~t$ to cst$(\mathrm{Pe})\sqrt{t}$  and, consequently, a Rouse-like subdiffusion would be observed for both $\mathrm{Pe}=0$ and $\mathrm{Pe}\neq0$. Were it in free space, the ABP would have transient superdiffusive dynamics before the final drift regime~\cite{wu,um2019}.  Since $B\sim 1/v_\mathrm{p}^2$ vanishes as $\mathrm{Pe}\to\infty$, the $\mathrm{ln}~t$ dynamics will be clearly manifested at high P{\'e}clet numbers. This is indeed demonstrated in Fig.~\ref{fig9} (Top), where the limiting curve $\frac{\tau_A\tau_R \pi}{4N^2}(\mathrm{ln}~t+C)$ is drawn as a dashed line for reference. When $\mathrm{Pe}$ is increased to $250$, the MSD up to the Rouse regime accurately follows this line.  

We can also extract the information about the apparent anomaly exponent $\alpha$ from Eq.~\eqref{eq:MSDSim}. As observed in Fig.~\ref{fig9}, in the Rouse regime the MSD grows like a power-law up to $\mathrm{Pe}\sim25$. We can reasonably estimate the apparent $\alpha$ as the minimum slope of the MSD in the log-log scale. Then, the minimum slope is identified to $\alpha= f'(u=u^*)$ where  $u={\rm ln}~t$ and $  f(u) = {\rm ln}\left(B{\rm e}^{u/2}+u+C\right)$. Here, $u^*$ is the time at the minimum slope satisfying $f''(u^*)=0$ (NB: the drift term can be ignored in $f$ because of $t/\tau_R\ll\sqrt{t/\tau_R}$ in the Rouse regime). 
From this method we find the apparent anomalous exponent $\alpha$ as a function of the P{\'e}clet number as below: 
\begin{equation}\label{eq:alpha}
\alpha(\mathrm{Pe}) = \frac{1}{2+2{\rm W} \left(\frac{2{\rm e^{C/2-2}}}{B}\right)}.
\end{equation}
Here, $\mathrm{W}(\cdot)$ is the Lambert W-function defined as the inverse of $\mathcal{F}(x)=x e^x $. In Fig.~\ref{fig9} (Bottom), we compare Eq.~\eqref{eq:alpha} to $\alpha(\mathrm{Pe})$ from the power-law fit to the simulation data (Fig.~\ref{fig9} (Top)). It shows that our theory with no free parameters successfully explains the observed apparent $\alpha$ up to $\mathrm{Pe}\sim 20$, as expected from the MSD curves. Beyond this value, the theory begins to disagree with the empirical values from a blind fit. This is because for $\mathrm{Pe} \gtrsim 20$ the effect of $\mathrm{ln}~t$ is increasingly pronounced compared to $B\sqrt{t}$, where the minimum point $u^*$ is located above the Rouse regime and, thus, the estimation \eqref{eq:alpha} is no longer applicable. 

\begin{figure}
	\centering
	\includegraphics[width=8.5cm]{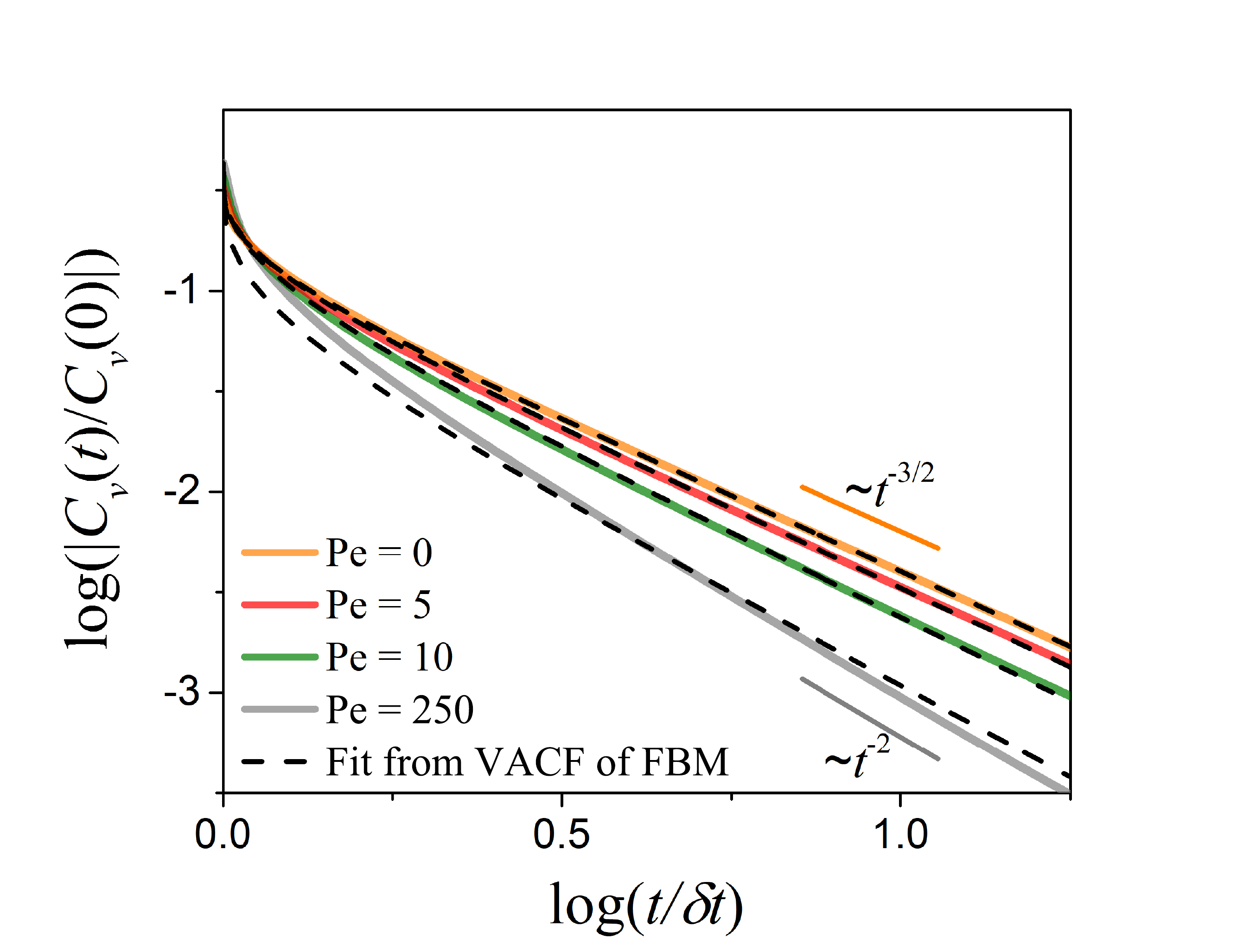}
	\caption{The theoretical VACF curves [Eq.~\eqref{eq:Cvtheory}] for the ABP cross-linker for various $\mathrm{Pe}$ conditions. The relaxation part for $t\geq \delta t$ is plotted and numerically fitted by the normalized VACF [Eq.~\eqref{eq:vacf_fbm}] of FBM (dashed line) with $H$ as a fit parameter. 
		From Top to Bottom: the fitted values of $H$ are $0.506$, $0.431$, $0.330$, and $0.175$.
	}
	\label{fig10}
\end{figure}

Now we shortly discuss about the VACF $C_v(t;\delta t)$ of the ABP cross-linker. Its analytic expression can be obtained using the autocorrelation relation [Eq.~\eqref{eq:Xp_autocorrelation}] with the entire derivation described in Sec.~II C (ESI\dag). The calculation shows that the VACF is given by the sum of two independent contributions, such that
\begin{equation}\label{eq:Cvtheory}
C_v(t;\delta t)=\frac{\tau_A\tau_R v_\mathrm{p}^2 \pi}{4N^2}\{BC_v^{(\mathrm{th})}(t)+C_v^{(\mathrm{ac})}(t)\}
\end{equation}  
where $B$ is the expansion coefficient defined in Eq.~\eqref{eq:MSDSim}. Fig.~S6 illustrates the excellent agreement between Eq.~\eqref{eq:Cvtheory} and the simulation data. 
Here, $C_v^{(\mathrm{th})}$ explains the displacement correlation from the thermal collective dynamics, which is, for a typical time scale (i.e., $\tau_A\ll t\ll \tau_R$ and $\delta t\ll\tau_R$), given by
\begin{eqnarray}\label{eq:Cvtheory2}
C_v^{(\mathrm{th})}(t)&=& \left[2 t^{1/2}-(t+\delta t)^{1/2}-(t-\delta t)^{1/2}\right]\propto -t^{-3/2}.\quad 
\end{eqnarray}
This is the precisely VACF [Eq.~\eqref{eq:vacf_fbm}] for FBM with $H=1/4$ characterized with a power-law tail with the exponent $-3/2$. The other term $C_v^{(\mathrm{ac})}$ is attributed to the displacement correlation from the active collective motion and, for the time scale of our interest, has a tail ($t> \delta t$)
\begin{eqnarray}\label{eq:Cvtheory3}
C_v^{(\mathrm{ac})}(t) &=& \left[{\rm ln}\left(t+\delta t\right)+{\rm ln}\left(t-\delta t\right)-2{\rm ln}~t \right] \\
&\propto& -t^{-2}.
\end{eqnarray}
Note that $C_v^{(\mathrm{ac})}\sim \partial^2{\rm ln}~t/\partial t^2$ has a power-law correlation with the exponent $-2$, which is the limiting power-law tail of FBM [Eq.~\eqref{eq:vacf_fbm}] at $H\to0$. The VACFs [Eqs.~\eqref{eq:Cvtheory}, \eqref{eq:Cvtheory2}, and \eqref{eq:Cvtheory3}] inform that the motion of the ABP cross-linker is a non-Markovian gaussian process with a power-law displacement correlation and seemingly behaves as FBM, particularly at small $\mathrm{Pe}$ where the contribution of $C_v^{(\mathrm{th})}$ relative to $C_v^{(\mathrm{ac})}$ is significant ($B\sim 161/\mathrm{Pe}^2$). In Fig.~\ref{fig10}, we plot the exact theoretical VACF [Eq.~\eqref{eq:Cvtheory}, solid line] for an increasing $\mathrm{Pe}$ and fit each curve with that of FBM [Eq.~\eqref{eq:vacf_fbm}, dashed line]. For $\mathrm{Pe}\lesssim 10$, the VACF is in good agreement with an FBM counterpart with a Hurst exponent from fitting. Hence, at a $\mathrm{Pe}$ condition in this range, the athermal motion of the ABP cross-linker can be effectively treated as FBM. For $\mathrm{Pe}\gtrsim 10$, there is no FBM curve fitting the VACF in the entire time range considered (e.g., the case of $\mathrm{Pe}=250$ in the plot). In this range of $\mathrm{Pe}$, $C_v^{(\mathrm{ac})}$ dominates the displacement correlation, and the dynamics of the ABP cross-linker are governed by another non-Markovian process having a power-law autocorrelation Eq.~\eqref{eq:Cvtheory3}. 

\section{Conclusions}
\label{sec:conclusion}

To summarize, in this work, we investigated the dynamics of an active Brownian particle cross-linked to a networked polymer based on the Langevin dynamics simulation and analytical treatment. It has been determined that the ABP cross-linker displays an apparent anomalous diffusion of the form $\langle\Delta \mathbf{R}^2(t)\rangle\sim t^\alpha$ with $0<\alpha\leq 1/2$ in the Rouse regime through the viscoelastic feedback. Contrary to naive expectations, the apparent anomaly exponent $\alpha$ becomes increasingly smaller than the Rouse exponent $1/2$ as the ABP is driven with a larger P{\'e}clet number $\mathrm{Pe}$. Thus, it appears that a particle with higher activity becomes more subdiffusive when the MSD is measured. Our analysis has also shown that the motion of ABP cross-linker is a non-Markovian gaussian process and seemingly behaves as FBM with the Hurst exponent given by $H=\alpha(\mathrm{Pe})/2$ in the Rouse regime. These dynamic properties are generic in the sense that they remain unaffected by the change of the functionality $f$ and the boundary conditions of the end beads in the network. 

We have provided an analytic theory regarding the ABP connected to a viscoelastic polymer environment. The analysis has demonstrated that in the Rouse regime the ABP dynamics are governed by a drift [$\sim t$], the thermally-induced collective (Rouse) [$\sim t^{1/2}$], and the actively-induced collective [$\sim \ln t$] dynamics. The ABP, through the viscoelastic feedback, attains strong negative spatial correlations of a power-law decay $-t^{-3/2}$ from the thermal noise and $ -t^{-2}$ from the active noise. It turns out that the apparent anomalous diffusion $\sim t^\alpha$ in the simulation is the superimposed dynamics of the three. The motion looks more subdiffusive for higher $\mathrm{Pe}$, as the contribution of the active dynamics ($\ln t$) is increasingly stronger than the Rouse dynamics as $\mathrm{Pe}$ is increased [see Eq.~\eqref{eq:MSDSim}]. We highlight that the viscoelastic feedback gives rise to the $\ln t$ dynamics against a local athermal force exerted by the ABP. The active diffusion of the ABP is easily hampered by the negative viscoelastic feedback from the polymer that decays as $\propto -t^{-2}$. Consistent with this view, we have found that the ABP-driven fluctuations are weakened rapidly along the chain. The corresponding contribution for MSD decreases with the bead index $s$ as $M_s^{(3)}\sim\mathrm{cst.}\left[\frac{\pi s}{2N}- {\rm \ln}\left(\frac{\sqrt{2}\pi s}{N}\right)\right]$, which decays much faster than a simple linear decay [see Fig.~\ref{fig8} (Bottom)]. 

Importantly, the knowledge obtained from our study leads us to establish an effective theory for the diffusion dynamics of an active particle in a viscoelastic medium. Given the fact that the ABP cross-linker is governed by a non-Markovian gaussian process characterized by two VACFs [Eqs.~\eqref{eq:Cvtheory2} \& \eqref{eq:Cvtheory3}] from two independent noise sources, we suggest that the following fractional Langevin equation exactly describes such a (one-dimensional) process $X(t)$:
\begin{equation}\label{eq:GLE}
\int_0^t {\mathcal{K}(t-\tau)}\dot{X}(\tau) {\rm d}\tau = \Gamma\left(1/2\right)\frac{d^{1/2} X}{dt^{1/2}}=\zeta_\mathrm{th}(t) + \zeta_\mathrm{ac}(t).\quad
\end{equation}
Here, $\frac{d^{1/2}}{dt^{1/2}}$ is the Caputo fractional derivative of order $1/2$, and the corresponding memory kernel is given by $\mathcal{K}(t)\sim t^{-1/2}$. The two random sources $\zeta_\mathrm{th}$ and $\zeta_\mathrm{ac}$, originating from the thermal and active noises in the microscopic theory [Eq.~\eqref{eq:method_abp}], respectively, incorporate the viscoelastic effect from the environment. As is known, $\zeta_\mathrm{th}$ is the  fractional gaussian noise (FGN) of $H=3/4$ satisfying FDT $\mathcal{K}(t-t')\propto \langle \zeta_\mathrm{th}(t)\zeta_\mathrm{th}(t')\rangle \sim |t-t'|^{2H-2}$~\cite{ralfpccp2,lizanaPRE_fbm,sakauePRE2013}.  
This then generates the Rouse dynamics $\langle \Delta X_\mathrm{th}^2(t)\rangle \sim t^{1/2}$ and the VACF \eqref{eq:Cvtheory2} for $X_\mathrm{th}(t)$. For the same kernel $\mathcal{K}(t)$, $\zeta_\mathrm{ac}$ should be an OU noise with an exponentially decaying autocorrelation $\langle \zeta_\mathrm{ac}(t)\zeta_\mathrm{ac}(t')\rangle\propto \exp(-|t-t'|/\tau_\mathrm{ac})$. Its amplitude and characteristic time are specified by the details of the microscopic active noise $\eta(t)$ and of the environment. In association with the $\mathcal{K}(t)$, $\zeta_\mathrm{ac}$ produces the MSD of $\langle \Delta X_\mathrm{ac}^2(t)\rangle\sim \ln t$ and the VACF decaying as $-t^{-2}$~\cite{GLE1996}, while violating FDT. Since the displacements $\Delta X_\mathrm{th}(t)$ and $\Delta X_\mathrm{ac}(t)$ are independent each other, the fractional Langevin equation \eqref{eq:GLE} is indeed the effective single-particle theory describing 
the observed anomalous dynamics of the ABP cross-linker. This new theoretical formalism will be useful for the quantitative understanding of the dynamics and physical properties of active particles in a viscoelastic environment, as exemplified in Fig.~\ref{fig1} (Top).    

We note that our findings in this work are not specific to the active noise we chose. Based on additional simulations, we have observed that a different type of active cross-linker, modeled by a white gaussian noise, produces essentially the same anomalous dynamics as above using the OU noise [Eq.~\eqref{eq:OUauto}]. The corresponding simulation data and theory are provided in Fig.~S7 and Sec.~II D (ESI\dag). Previously, the active-particle models using these white gaussian nonequilibrium forces were introduced for studying the \textit{in vivo} chromosome \cite{menon2014_NAR,smrek2017_PRL} and other active systems \cite{leticia2011_softmatter,smrek2020_natcomm,leticia2008_PRE,luijten2017_PNAS}, with the notion of "hot" and "cold" particles~\cite{menon2014_NAR,smrek2020_natcomm}. The superimposition of the active gaussian white noise with the thermal noise makes the active particles in an effective temperature hotter than the ambient temperature that the passive particles experience. It is likely that our findings in this study can be observed in the dynamics of other FDT-violating particles coupled to a polymeric environment. 

Finally, it is worth mentioning that the current study potentially provides a basic knowledge for understanding the nonequilibrium dynamics of the biological polymer complexes, such as chromosome, cytoskeleton filaments, and the ER networks, as well as for interpreting the related experimental data. For example, imagine a fluorescence in situ hybridization (FISH) experiment probing local segment dynamics in a living polymer associated with the ATP-consuming enzymatic activity. Our computational and analytic investigations suggest that \textit{active} subdiffusive motion can emerge through the interactions between the active particles and the underlying polymer, with an anomaly exponent $\alpha$ lower than the one expected at the thermal equilibrium, e.g., $1/2$ for the case of a Rouse polymer. Note that the active sudiffusion shown in this work should be distinguished from the athermal subdiffusion of a self-propelled particle or an active polymer that was induced by the macromolecular crowding of the environment~\cite{weissPRE_ER,weiss2017_NJP}. Therefore, a careful interpretation is necessary for the experimental observation, and one may tell the active subdiffusion from the ordinary case by scrutinizing the decrease in $\alpha$ with increasing active fluctuations (say, by increasing the ATP concentration in the solution or increasing a local temperature at the probed region via chemical reaction).    


\section*{Conflicts of interest }
There are no conflicts to declare.  

\section*{Acknowledgements}
This work was supported by the National Research Foundation (NRF) of S. Korea through No.~2018R1A6A3A11043366 (O.L) and No.~2020R1A2C4002490 (J-H.J). We are grateful to Won Kyu Kim and Ludvig Lizana for the valuable feedback on the manuscript. J-H.J also acknowledges the hospitality and academic support from the Korea Institute for Advanced Study. 



\balance


\bibliographystyle{main_PRX} 
\bibliography{main_PRX} 

\end{document}